\documentclass[num-refs]{nbdt-article}

\usepackage{siunitx}

\pdfoutput=1

\papertype{Original Article}
\paperfield{Behavior}

\title{Feedback Gains modulate with Motor Memory Uncertainty}

\abbrevs{EMG, electromyography; CW, clockwise; CCW, counterclockwise; MPE, maximum perpendicular error; ANOVA, analysis of variance.}

\author[1]{Sae Franklin}
\author[2]{David W. Franklin}

\affil[1]{Institute for Cognitive Systems, Department of Electrical and Computer Engineering, Technical University of Munich, Germany}
\affil[2]{Neuromuscular Diagnostics, Department of Sport and Health Sciences, Technical University of Munich, Germany}

\corraddress{David W. Franklin, Neuromuscular Diagnostics, Department of Sport and Health Sciences, Technical University of Munich, Germany}
\corremail{david.franklin@tum.de}

\runningauthor{Franklin and Franklin}

\begin{document}

\maketitle

\begin{abstract}
A sudden change in dynamics produces large errors leading to increases in muscle co-contraction and feedback gains during early adaptation. We previously proposed that internal  model  uncertainty drives  these  changes,  whereby  the sensorimotor system reacts to the change in dynamics by up regulating stiffness and feedback gains to reduce the effect of model errors. However, these feedback gain increases have also been suggested to represent part of the adaptation mechanism. Here, we investigate this by examining changes in visuomotor feedback gains during gradual or abrupt force field adaptation. Participants grasped a robotic manipulandum and reached while a  curl  force  field  was  introduced  gradually  or  abruptly. Abrupt introduction of dynamics elicited large initial increases in kinematic error, muscle co-contraction and visuomotor feedback gains, while gradual introduction showed little initial change in these measures despite evidence of adaptation. After adaptation had plateaued,there was a change in the co-contraction and visuomotor feedback gains relative to null field movements, but no differences (apart from the final muscle activation pattern) between the abrupt and gradual introduction of dynamics. This suggests that  the  initial  increase  in  feedback  gains  is  not  part  of  the  adaptation process, but instead an automatic reactive response to internal model uncertainty. In contrast, the final level of feedback gains is a predictive tuning of the feedback gains to  the  external  dynamics as  part of  the  internal  model adaptation. Together, the reactive and  predictive  feedback  gains  explain  the wide variety  of previous experimental results of feedback changes during adaptation.

\end{abstract}

\section{Introduction}
Humans have exceptional abilities to skillfully manipulate and interact with objects in the environment. Our sensorimotor system constantly generates appropriate signals to control our musculoskeletal system, based on a prediction of the current dynamics. When these dynamics are stable and repeatable, we use an internal model or motor memory to produce an efficient and effective motion. When the external dynamics change, our sensorimotor control system adapts rapidly to the environmental disturbances in a manner that indicates a fundamental knowledge of the mechanics of the external world \cite{lackner_rapid_1994, shadmehr_adaptive_1994, conditt_motor_1997}. Any sudden change in the environmental dynamics during a movement causes large kinematic errors, leading to a rapid increase in muscle co-contraction \cite{thoroughman_electromyographic_1999, osu_short-_2002, franklin_adaptation_2003, milner_impedance_2005}. This co-contraction increases limb stiffness, both intrinsic \cite{hoffer_regulation_1981, kearney_identification_1997, mirbagheri_intrinsic_2000} and reflexive contributions \cite{akazawa_modulation_1983, saliba_co-contraction_2020}, and acts to limit the perturbing effects of the dynamics until the sensorimotor control system is able to learn a motor memory or internal model that can predictively compensate for these dynamics. Once this motor memory or internal model is updated, this co-contraction is gradually decreased.

Large kinematic errors which occur during the initial stages of adaptation to novel dynamics have also been shown to produce rapid increases in visuomotor \cite{franklin_visuomotor_2012} and long latency stretch \cite{coltman_time_2020} feedback responses. We previously proposed that this increase resulted from uncertainty in the internal model \cite{franklin_visuomotor_2012}. During initial adaptation, the sensorimotor system receives unexpected error signals indicating that our current internal model no longer predicts the environment accurately, and that it either needs to update our current motor memory or select a new motor memory \cite{oh_minimizing_2019}. We suggest that along with co-contraction, the sensorimotor control system also upregulates feedback gains until the sensorimotor control system is able to relearn a new model that predictively compensates for these dynamics. These feedback gains are a reactive response to the internal model uncertainty to limit the perturbing effect of the novel dynamics during this initial phase of adaptation. However, even after adaptation to the new environment, these feedback gains are upregulated compared to the null field levels and tuned to the environmental dynamics \cite{franklin_visuomotor_2012, franklin_rapid_2017, cluff_rapid_2013}. That is, these feedback gains at the end of the adaptation process appear to arise as the sensorimotor control system regulates the gain of the feedback system as part of the adaptation process to novel dynamics, resulting in this final predictive component of feedback gains. We therefore suggested that there are two computational components of increased feedback gains; a reactive response to model uncertainty and a predictive response that is learned as part of the internal model \cite{franklin_rapid_2017}.

Here we examine whether the initial reactive component of the increased feedback gain is driven by internal model uncertainty or is simply learned as part of the adaptation process by contrasting the adaptation to abrupt changes in dynamics with the gradual introduction \cite{malfait_is_2004, klassen_learning_2005, kluzik_reach_2008, huang_persistence_2009, orban_de_xivry_contributions_2011, pekny_protection_2011, milner_different_2018}. An abrupt introduction to a novel force field causes large error signals, whereas a gradual introduction to the same force field produces little or no error signals. Large errors would indicate that our current internal model no longer well predicts the environmental dynamics. In contrast, a gradual introduction of the force field provides only small errors within the natural variability of human reaching movements. As such, we predict that there will be little change in the uncertainty associated with the internal model, thereby causing little or no reactive increase in the feedback gains, despite adaptation continuing throughout the exposure phase. Instead, we predict only a gradual increase in the visuomotor feedback gains as the appropriate level of steady state feedback gains is learned for the dynamics. Here, we test these predictions by having participants adapt to both gradual and abrupt dynamics while measuring their visuomotor feedback gains to visual motion of the hand \cite{brenner_fast_2003, sarlegna_target_2003, saunders_humans_2003, franklin_specificity_2008, knill_flexible_2011}. Understanding the role that errors, rate of environmental change and internal model uncertainty have on the adaptation process, especially on co-contraction and feedback gains, provides us with important knowledge about structuring learning in sports, rehabilitation and other motor skills. 

\section{Material and Methods}
\subsection{Experimental Setup}
\subsubsection{Participants}
Twelve participants participated in the experiment (3 male and 9 female: aged 25.0 $\pm$ 4.4, mean $\pm$ SD). All participants were right-handed according to the Edinburgh handedness inventory \cite{oldfield_assessment_1971} with no reported neurological disorders. Participants provided written informed consent, and the institutional ethics committee approved the experiments. One additional participant took part in the experiment, but was excluded from the final analysis as the level of force field adaptation was 34.7\% across both force fields (CW and CCW) at the end of the exposure period. All other participants achieved a mean above 60\% across the abrupt and gradual adaptation protocols. The exclusion of this participant as an outlier was confirmed using Grubb’s test across the final fifteen blocks of exposure in both abrupt and gradual conditions.
\subsubsection{Apparatus}
Participants grasped the handle of the vBOT robotic manipulandum \cite{howard_modular_2009} with their forearm supported against gravity with an air sled (Fig. 1A). The robotic manipulandum both generated the environmental dynamics (null field, force field or channel), and measured the participants’ behavior. Position and force data were sampled at 1KHz. Endpoint forces at the handle were measured using an ATI Nano 25 6-axis force-torque transducer (ATI Industrial Automation, NC, USA). The position of the vBOT handle was calculated from joint-position sensors (58SA; IED) on the motor axes. Visual feedback was provided using a computer monitor mounted above the vBOT and projected veridically to the participant via a mirror. This virtual reality system covers the manipulandum, arm and hand of the participant, preventing any visual information about their location. The exact time that the stimuli were presented visually to the participants was determined using the video card refresh rate and confirmed with an optical sensor to prevent a time delay. Participants performed right-handed forward reaching movements in the horizontal plane at approximately 10 cm below shoulder level.

\begin{figure}[t]
\centering
\includegraphics[width=12cm]{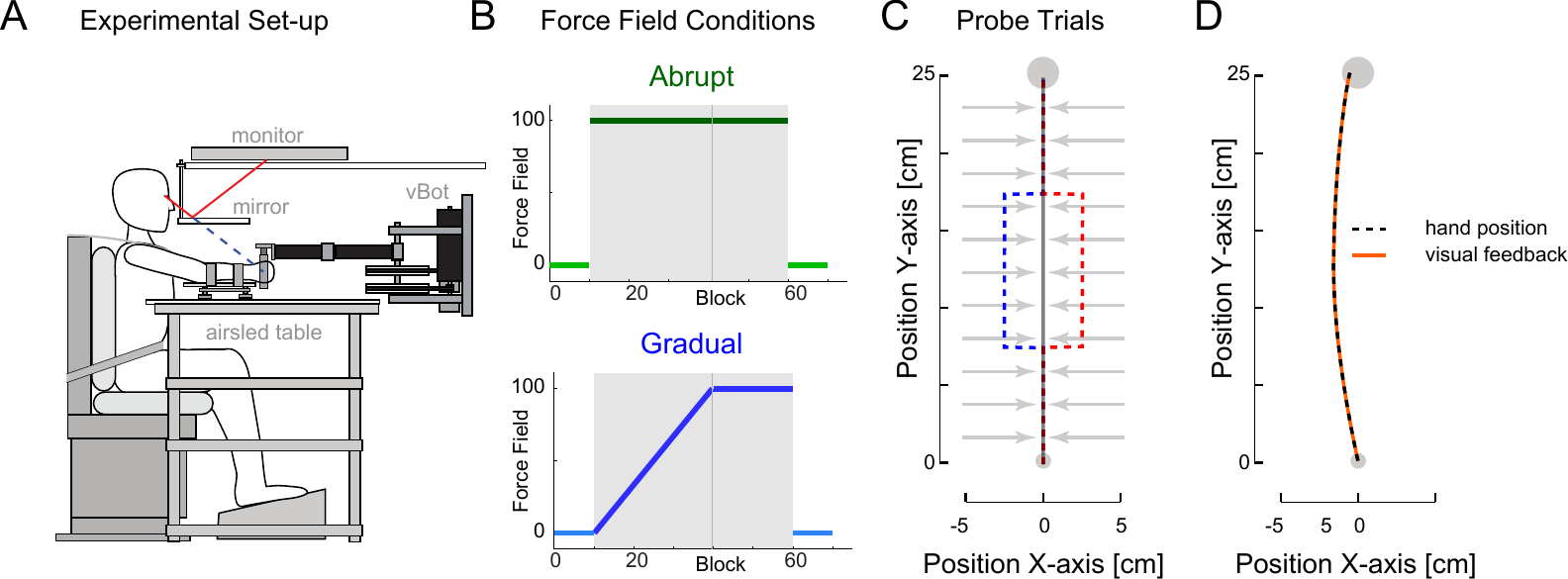}
\caption{Experimental  set-up.  A:  Participants  grasp  the  robotic  manipulandum (vBOT)  while  seated  and  visual  feedback  is  presented  veridically  using  a  top mounted  monitor  viewed  through  a  mirror  so  that  it  appears  in the  plane  of movement.  The  participant’s  forearm  is  supported  by  an  airsled.  B:  Participants experienced either an abrupt (green) or gradual (blue) onset of a velocity dependent force  field  in  different  sessions.  C:  Throughout  the  experiment,  on  random  trials visual perturbations (probe trials) were used to examine the magnitude of the visually induced motor response. During these trials, the hand (dark grey line) was physically constrained  to  the  straight  path  between  the  initial  starting position and  the final target using a mechanical channel, such that any force produced in response to the visual perturbation (blue or red dotted lines) can be measured against the virtual channel wall (grey arrows) using the force sensor. D: In all trials except the probe trials, the hand cursor (orange line) is shown to the participant overlaid on the actual hand movement (black dotted line).}
\end{figure}

\subsubsection{Setup}
Participants were seated with their shoulders restrained against the back of a chair by a shoulder harness. Movements were made from a 1.0 cm diameter start circle centered approximately 28.0 cm in front of the participant to a 2.0 cm diameter target circle centered 25 cm in front of the start circle. The participant’s arm was hidden from view by the virtual reality visual system, on which the start and target circles as well as a 0.6 cm diameter cursor used to track instantaneous hand position were projected. Participants were instructed to perform successful movements to complete the experiment. A successful movement required the hand cursor to enter the target (without overshooting) within 700 $\pm$ 75 ms of movement initiation. Overshoot was defined as movements that exceeded the target in the direction of movement. When participants performed a successful movement they were provided with feedback as how close they were to the ideal movement time of 700ms (‘great’ or ‘good’) and a counter increased by one. Similarly, when they performed unsuccessful movements they were provided with feedback as to why the movement was not considered successful (“too fast”, “too slow” or “overshot target”). Trials were self-paced; participants initiated a trial by moving the hand cursor into the start circle and holding it within the target for 1000 ms. A beep then indicated that the participants could begin the movement to the target. The duration of the movement was determined from the time that the participants exited the starting position until the time that participants entered the target.
\subsubsection{Electromyography}
Surface electromyography (EMG) was recorded from two mono-articular shoulder muscles (pectoralis major and posterior deltoid), two bi-articular muscles (biceps brachii and long head of the triceps) and two mono-articular elbow muscles (brachioradialis, and lateral head of the triceps). The EMG was recorded using the Delsys Bagnoli (DE-2.1 Single Differential Electrodes) electromyography system (Boston, MA). The electrode locations were chosen to maximize the signal from a particular muscle while avoiding cross-talk from other muscles. The skin was cleaned with alcohol and prepared by rubbing an abrasive gel into the skin. This was removed with a dry cotton pad and the gelled electrodes were secured to the skin using double-sided tape. The EMG signals were analog band-pass filtered between 20 and 450 Hz (in the Delsys Bagnoli EMG system) and then sampled at 2.0 kHz. 

\subsubsection{Probe trials to measure reflex gain}
In order to assess reflex magnitude, visually induced motor responses were examined using perturbations of the visual system similar to those previously described \cite{franklin_specificity_2008, franklin_visuomotor_2012, franklin_fractionation_2014, dimitriou_temporal_2013, reichenbach_dedicated_2014} throughout the experiments. On random trials, in the middle of a movement to the target, the cursor representing the hand position was jumped perpendicular to the direction of the movement (either to the left or to the right) by 2 cm for 250 ms and then returned to the true hand position for the rest of the movement (Fig. 1C). During these trials, the hand was physically constrained to the straight path between the starting position and the target using a mechanical channel, such that any force produced in response to the visual perturbation could be measured against the channel wall using the force sensor. The mechanical wall of the channel was implemented as a stiffness of 5,000 N/m and damping of 2Nm\begin{math}^{-1} \end{math}s for any movement lateral to the straight line joining the starting location and the middle of the target \cite{scheidt_persistence_2000, milner_impedance_2005}. As this visual perturbation was transitory, returning to the actual hand trajectory, participants were not required to respond to this visual perturbation to produce a successful trial. These visual perturbations were applied perpendicular to the direction of the movement (either to the left or the right). For comparison, a zero-perturbation trial was also included in which the hand was held to a straight-line trajectory to the target, but the visual cursor remained at the hand position throughout the trial. The onset of the displacements occurred starting at 7.5 cm (30\% of the length of the movement). The perturbation trials were randomly applied during movements in a blocked fashion such that one of each of the three perturbations were applied within a block of twelve trials.

\subsection{Experimental Paradigm}
Our previous experiment suggested that the large feedback gains during initial adaptation were due to the uncertainty in the internal model \cite{franklin_visuomotor_2012}, however this could not be dissociated from motor adaptation. Here we examine this phenomenon in detail, by contrasting the adaptation to abrupt changes in dynamics with the gradual introduction \cite{kluzik_reach_2008, orban_de_xivry_contributions_2011}. 
In this experiment, we examine the changes in the feedback gains while the reaching errors are presented both physically and visually to participants. While participants could not see their hands, a visual cursor (yellow circle) was presented which corresponded to their hand location. With the exception of 250ms on the probe trials in which this cursor was perturbed laterally by 2cm, the cursor always matched the physical location of their hand in both x- and y-axes (Fig. 1D). 

\subsubsection{Experimental Protocol}
Each participant performed two sessions, which were separated by a short break, in a single day. On one session, a dynamical force field was abruptly introduced (abrupt), while on the other session a directionally opposite force field was gradually applied (gradual). The order of both sessions and the force field directions were counterbalanced across participants. EMG electrodes remained in place throughout the entire experiment. Before each session, participants performed a practice of 61 null field trials in order to familiarize themselves with the movement criteria. Throughout the experiment, trials were arranged in blocks consisting of twelve trials; of which three were probe trials (visual perturbation and mechanical channel) and nine were normal reaching trials (null field or force field depending on the phase of the experiment). These probe trials were used to assess both the visuomotor feedback gain and the degree of learned force compensation. While lateral movement in the random probe trials was constrained by the mechanical channel, participants were free to move in any direction during all other trials.
Each session consisted of 4 phases. A single movement was always performed first in any new phase such that a probe trial was never the first movement. First, participants experienced the pre-exposure phase of 121 null field trials (10 blocks of twelve trials plus one initial trial). Next, the initial exposure stage consisted of 361 force field trials (30 blocks plus one initial trial). In the abrupt condition, the full force field was applied from the first trial, whereas in the gradual condition the force field was scaled up from one trial to the next over the 361 trials (Fig. 1B). The force field was a velocity-dependent curl force field where the force in N (Fx, Fy) on the hand was computed as:
\begin{eqnarray}
\left[
\begin{array}{ccc} F_{x} \\ F_{y} \end{array}\right]=b\left[
\begin{array}{ccc}
0 & -1 \\
1 & 0 \\
\end{array}\
\right]\left[
\begin{array}{ccc}
\Dot{x} \\
\Dot{y}\\
\end{array}
\right]
\end{eqnarray}
depending on the participants' hand velocity ($\Dot{x},\Dot{y}$) [m/s] and the scaling factor $b$ which was either 0.16 N/m/s (clockwise curl force field: CW) or -0.16 N/m/s (counter-clockwise curl force field: CCW). Once the initial exposure phase was completed, both groups of participants performed the final exposure phase consisting of another 20 blocks of trials (241 trials). Finally, participants experienced the post-exposure phase in which 10 blocks (121 trials) of null field trials were performed. Participants were required to take short breaks every 200 movements throughout the experiment. They were also allowed to rest at any point they wished by releasing a safety switch on the handle. 
\subsection{Analysis}
Analysis of the experimental data was performed using Matlab R2019a. EMG data were band-pass filtered (30 – 500Hz) with a tenth-order, zero phase-lag Butterworth filter and then rectified. Position, velocity and endpoint force were low-pass filtered at 40 Hz with a fifth-order, zero phase-lag Butterworth filter. Statistics were performed in Matlab and JASP 0.14.1 \cite{JASP2020}. Statistical significance was considered at the p<0.05 level for all statistical tests.
\subsubsection{Hand Path Error}
The maximum perpendicular error (MPE) was used as a measure of the straightness of the hand trajectory. On each trial, the MPE is the maximum distance on the actual trajectory that the hand reaches perpendicular to the straight-line path joining the start and end targets (errors to the left are defined as negative and errors to the right are defined as positive). The MPE was calculated for each non-probe trial throughout the learning experiment.
\subsubsection{Force Compensation}
In order to examine the predictive forces exerted by the participants throughout the experiment, the forces against the channel walls on the probe trials were used. On each trial, the amount of force field compensation was calculated by linear regression of the measured lateral force against the channel wall onto the ideal force profile required for full force field compensation \cite{smith_interacting_2006}. Specifically, the slope of the linear regression through zero is used as a measure of force compensation. The ideal force field compensation was estimated as the product of the y-velocity and the force field scaling factor. In the null field, the ideal force field compensation is based upon the compensation required in the curl force field \cite{howard_gone_2012}. Therefore, values in the null force field before learning (pre-exposure phase) should be close to zero. In the gradual curl force field, the force compensation on a given trial was calculated based on the current strength of the force field at this specific trial.
\subsubsection{Electromyographic Activity}
For plotting purposes, the EMG was adjusted to the mean value of EMG in the null field trials prior to the force field exposure for each muscle of each participant prior to averaging. To examine differences in the overall muscle activity across the experiment, the integral of the rectified EMG data was taken over 700 ms from 100 ms before movement start until 600 ms after movement start. The EMG data was averaged across all trials in a block. The EMG was normalized and then averaged across participants (but not across muscles). To normalize for a particular muscle, a single scalar was calculated for each participant and used to scale the muscle’s EMG traces for all trials for that participant. The scalar was chosen so that the mean (across trials) of the EMG data averaged over the whole experiment was equal across participants (and set to the mean over all the participants). This puts each participant on an equal scale to influence any response seen in the data. For comparison across muscles, the EMG values were further scaled for each muscle relative to the mean value in the pre-exposure phase (expressed as a percentage value relative to the pre-exposure activity). The muscle activity was further separated into the amount related to co-contraction and the amount related to force production for each of the three muscle pairs. In order to examine the amount of co-contraction, the minimum value of EMG between the two muscles making up a muscle pair was determined and multiplied by 2 as both muscles would contribute to the increased stiffness. The activation that would correspond to the change in force was determined as the maximum muscle activity of the two muscles of the muscle pair subtracted by the minimum of the two muscle activities. Differences in these measures across the gradual and abrupt conditions were examined using a t-test in Matlab.
\subsubsection{Rapid Visuomotor Responses}
Individual probe trials were aligned on visual perturbation onset. The response to the right visual perturbation on probe trials was subtracted from the response to the left perturbation on probe trials in order to provide a single estimate of the motor response to the visual perturbation for each block. To examine the feedback gain, we calculated the average post-perturbation force over two intervals: the first corresponding to an rapid involuntary response (180–230 ms) \cite{franklin_specificity_2008}, and the second to a slower response (230-300 ms) \cite{franklin_fractionation_2014, franklin_rapid_2017}.
To examine the electromyographic responses to the visuomotor perturbations, the EMG traces were divided by the mean value of the muscle activity in the null force field (pre-exposure) for that muscle in that participant between -50 and 50 ms relative to the onset of the perturbation. Muscular responses were considered over two intervals: 90-120 ms and 120-180 ms as in previous studies \cite{dimitriou_temporal_2013, franklin_fractionation_2014, gu_trial-by-trial_2016, cross_visual_2019}.
\subsubsection{Comparisons}
In order to compare the final values after learning, the mean measures (MPE, force compensation and visuomotor feedback gain) over the last 15 blocks in the final exposure phase were contrasted between abrupt and gradual conditions using frequentist repeated measures ANOVAs with a between subjects factor of condition order using JASP 0.14.1. To complement the frequentist approach we also perform equivalent Bayesian repeated measures ANOVAs using JASP 0.14.1. Here we use the Bayes Factors $(BF_{10}$) to evaluate evidence for the null hypothesis. In order to examine the after effects on the first block of trials in the post-exposure phase for both MPE and force compensation, we subtracted the mean values in the pre-exposure phase from the value of the first block in the post-exposure phase. We then performed a paired t-test using Matlab to examine any differences between the gradual and abrupt adaptation groups. Significant differences were examined at the p<0.05 level.

\section{Results}
Participants performed forward reaching movements while grasping the handle of a robotic manipulandum (Fig. 1A). Participants were then presented with either an abrupt or gradual introduction of a velocity-dependent force field (Fig. 1B). Throughout the entire experiment we measured the feedback gains on random trials by presenting the participants with a brief visual perturbation of a hand cursor (to the right or left of the movement) while the physical hand was mechanically constrained to move within a channel to the target, termed probe trials, (Fig. 1C). On these probe trials we measured the lateral force produced by the participant’s hand into the channel providing information both about the learned predictive forces for compensating the force field and the magnitude of the rapid visuomotor feedback gains. Throughout the experiment, participants received visual feedback about their movement and the target via a computer screen in the plane of movement (Fig. 1D).
\subsection{Behavior}
All participants adapted to both an abrupt and a gradual introduction of a curl force field where both the order and the field direction were counterbalanced across participants. 
After a short pre-exposure phase where the initial trials were in the null field, a curl force field was applied unexpectedly. In the abrupt condition this caused a large increase in the kinematic error (Fig. 2A left, green). However, in the gradual condition, as the strength of the curl force field is gradually increased, the kinematic error (MPE) only slowly increased until the strength of the force field got close to the final level (Fig. 2A right, blue). Once the force field reached the final level, the MPE appeared to gradually reduce over the next twenty blocks of trials. Interestingly, after the experiment had finished, participants generally reported that they did not realize that a force field had been applied, instead only that something different was being done to their arm during these movements once they were halfway through the gradual loading of the force field. Comparing with a repeated measures ANOVA the last 15 blocks in the final exposure phase (dark gray shaded area), the MPE in the abrupt condition was not significantly different from that in the gradual condition ($F_{1, 10}=0.598, p=0.223; BF_{10}=0.845$) (Fig. 2A center bar graph) and there was no between subjects effect of condition order ($F_{1,10}=0.984, p=0.345; BF_{10}=0.604$). For both groups, when the force field was removed (post-exposure phase) the maximum perpendicular error increased dramatically in the opposite direction to that of the force field (Fig. 2A, left and right), but was reduced quickly over the subsequent blocks. There was no significant difference between the kinematic after effects on the first block of trials between the abrupt and gradual adaptation groups ($t_{11}=1.17, p=0.267$).

\begin{figure}
\centering
\includegraphics[width=11cm]{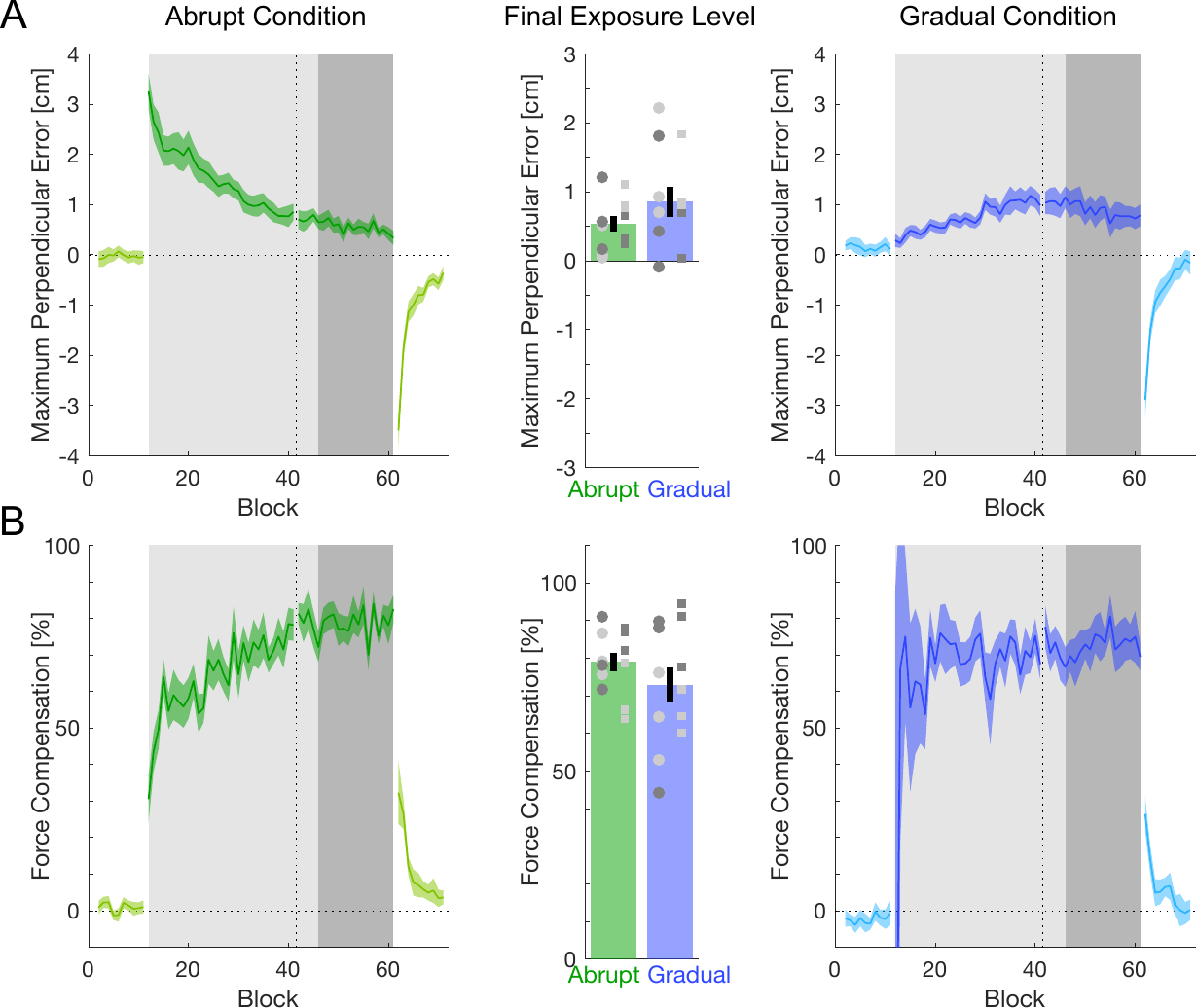}
\caption{Comparison of adaptation between abrupt and gradual exposure to the curl force fields. A: The mean (solid line) and standard error of the mean (shaded region) of the maximum perpendicular error (MPE) of the hand trajectory over the experiment. The sign of the MPE measurement from the CCW force fields was flipped so that all errors produced by the force field were shown to be positive. The gray shaded area (and darker green and blue colors) indicate the period over which the curl force field was applied. The vertical dotted line shows the time point at which the gradual force field was the same strength as the abrupt force field. Dark gray shaded area indicates the last 15 blocks of the full force field exposure phase over which the final levels of adaptation were compared. The center bar graph compares this final level of MPE across the two force fields. The error bar represents the standard error of the mean (s.e.m.). Each participant’s final value is shown with a point. A light grey point indicates the CW force field whereas a dark grey point indicates the CCW force field. The square indicates the first force field experienced whereas the circle indicates the second time participants experienced a force field. There were no significant differences between the conditions. B: Force compensation level over the experiment as measured on the channel trials. A value of 100\% indicates perfect compensation for the force field. Force compensation in the null field was quantified with respect to the full force field value, so a value of 0 is expected. In the gradual condition, as the force field is ramped up, the force compensation is expressed as a percentage of the force field that has been applied to the participant at this point in the experiment. Values plotted as in A. Similar levels of force compensation were found for abrupt and gradual conditions with no significant difference over the last 15 blocks of exposure.}
\end{figure}

Throughout the entire experiment, random trials were introduced in which participants performed movements in a mechanical channel constraining their hand to a straight movement to the target. Using these trials, we can estimate the predictive force compensation as participants adapt to the dynamics (Fig. 2B) indicating the percentage of perfect adaptation to the force field. Note that in the gradual condition, the percentage of adaptation is expressed as a function of the current level of force field during the ramp phase. Interestingly, the force compensation in the gradual force field adaptation appears to start from 60-70\% perfect adaptation from the beginning and stay around this level throughout both the initial ramping phase and the later constant phase. In both the abrupt and gradual conditions, participants showed high levels of force compensation (over 70\%) after they reached to 100\% strength of the force field (blocks 40-60). When comparing the final levels of adaptation in the abrupt and gradual conditions (final 15 blocks of exposure phase), we again found no significant difference (dark gray shaded area in the figures) in the force compensation (Fig. 2B center) between the abrupt and the gradual conditions ($F_{1,10}=0.962, p=0.350; BF_{10}=0.647$) and no between subjects effect of condition order ($F_{1,10}=1.280, p=0.284; BF_{10}=0.536$). Therefore, the final adaptation level of participants in the abrupt and gradual conditions were similar, although the abrupt condition had a much greater number of trials in which they were presented with the full magnitude of the force field. In the post-exposure phase, there was also no significant difference between the force compensation after effects on the first block of trials between the abrupt and gradual adaptation groups ($t_{11}=0.276, p=0.788$).

\subsection{Muscle Activity}
Muscle activity (EMG) was recorded from six muscles and is shown separately for the counter-clockwise (Fig. 3A) and clockwise (Fig. 3B) force fields. The EMG values are normalized to the null field level for all muscles, and are only shown here on the mechanical channel trials in which the force field is not applied and no trajectory errors occur. In the abrupt condition (green), all muscles increase their activity when the force field was first applied, suggesting an initial large increase in co-contraction. However, the muscles slowly reduced their activity as learning proceeded, eventually reaching a plateau level late in adaptation. At the end of the exposure phase there are high activity levels in the muscles compensating for the force field: posterior deltoid and triceps longus for the CCW force field and pectoralis major and biceps brachii for the CW force field. In the gradual condition (blue), the muscle activity remained low throughout the early exposure period, only gradually increasing and reaching a plateau towards the end of the exposure phase. The final levels of muscle activity were similar between both the abrupt and gradual conditions for both force fields. The similar levels of final muscle activity were true for both the muscles compensating for the force field, and those acting as stabilizers – increasing the co-contraction. When the force field was suddenly removed (after block 60), there were further increases of muscle activity, especially in the antagonist muscles – the muscles that were not acting to compensate for the force fields (e.g. posterior deltoid and triceps longus in the CW force field for a forward movement). As the muscle activity on this (and subsequent figures) is obtained from the channel trials, we would expect that most of the activity is pre-planned and not a reaction to kinematic errors experienced on the specific trial.

\begin{figure}[t]
\centering
\includegraphics[width=12cm]{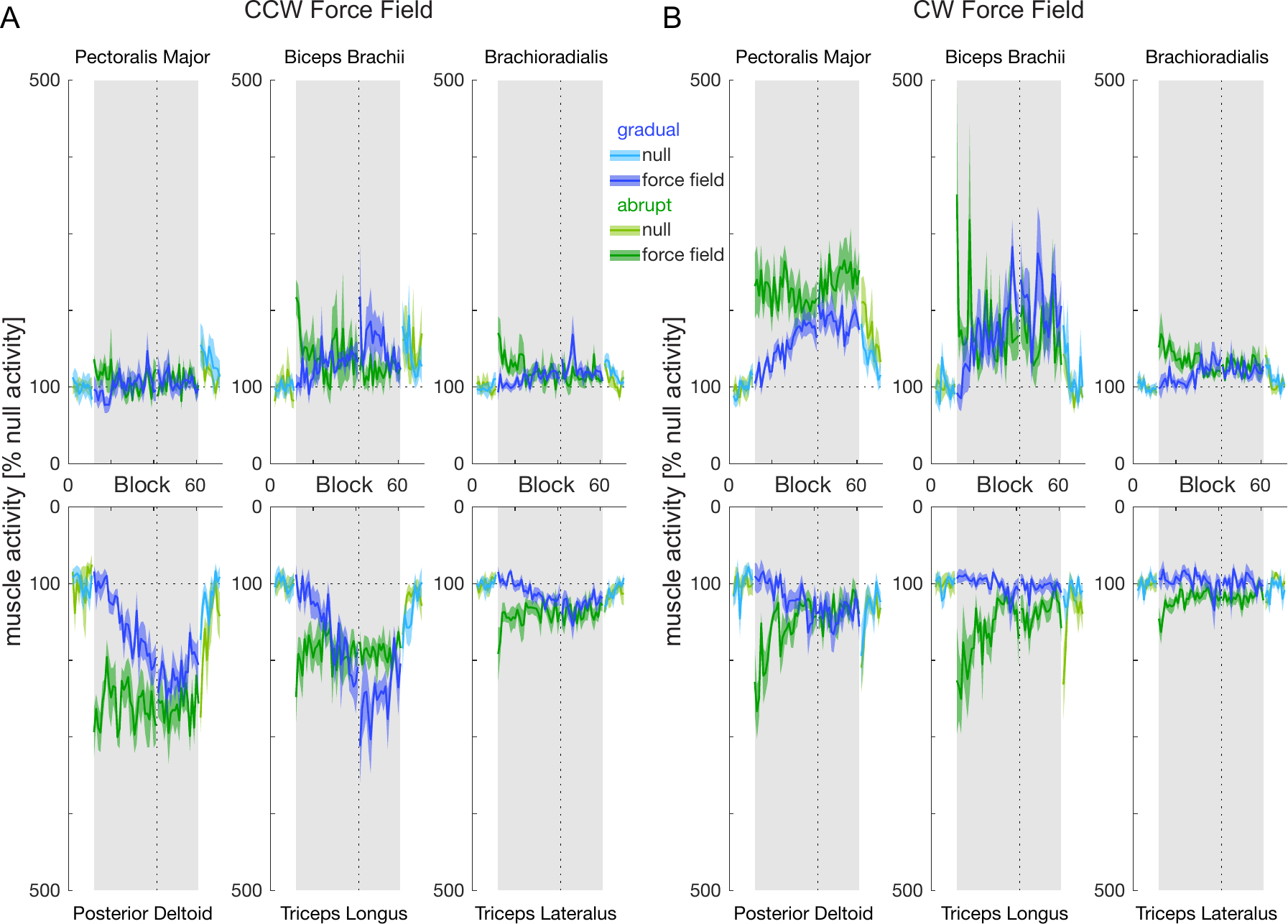}
\caption{Comparison of muscle activity (EMG) across the experiment for abrupt (green) and gradual (blue) exposure to curl force fields. Light blue and green indicate values in the null field whereas dark blue and green indicate values in the curl force field. Data is only from the randomly interspersed mechanical channel trials where the curl force field was not applied. Increasing extensor muscle activity is plotted in the downward direction. A: Muscle activity (mean and s.e.m) during adaptation to the CCW curl force field. Muscle activity was normalized for each of the six muscles to the mean value in the pre-exposure phase before averaging across participants. EMG values were calculated as the integrated muscle activity from -100 to 600 ms relative to the start of the movement. Grey shaded region indicates the exposure phase and the vertical dotted line indicates the point at which the gradual force field is equal to the full force field value. B: EMG in the CW curl force field. Across both force fields, high levels of muscle activation were initially observed when the force field was introduced abruptly (green) but not when introduced gradually (blue).}
\end{figure}

The temporal profile of muscle activity after adaptation was examined for both the CCW and CW force fields (Fig. 4) on the channel trials, with the profile of activity on the null field trials shown for comparison. As expected the profile of muscle activity in the null field prior to either the abrupt or gradual force field application is similar. After adaptation, we also find similar temporal profiles of muscle activation for many of the muscles, but with specific differences between the conditions. In particular, in the CCW force field (Fig. 4A), there was a larger activation of the posterior deltoid muscle in the abrupt condition, whereas there was a larger triceps longus activation in the gradual condition. After adaptation to the CW force field (Fig. 4B) it appears that there is a larger pectoralis major and triceps longus muscle activation after the abrupt introduction of the force field, whereas the gradual condition shows a slightly larger biceps brachii activation. Therefore, across both force fields, the abrupt introduction of the force field produced slightly higher activation of the shoulder muscles, whereas gradual adaptation often recruited the biarticular muscles to a larger degree. It is therefore possible that the specific pattern of large directional errors in the abrupt conditions changes the overall recruitment pattern of muscles even after adaptation.

\begin{figure}[t]
\centering
\includegraphics[width=12cm]{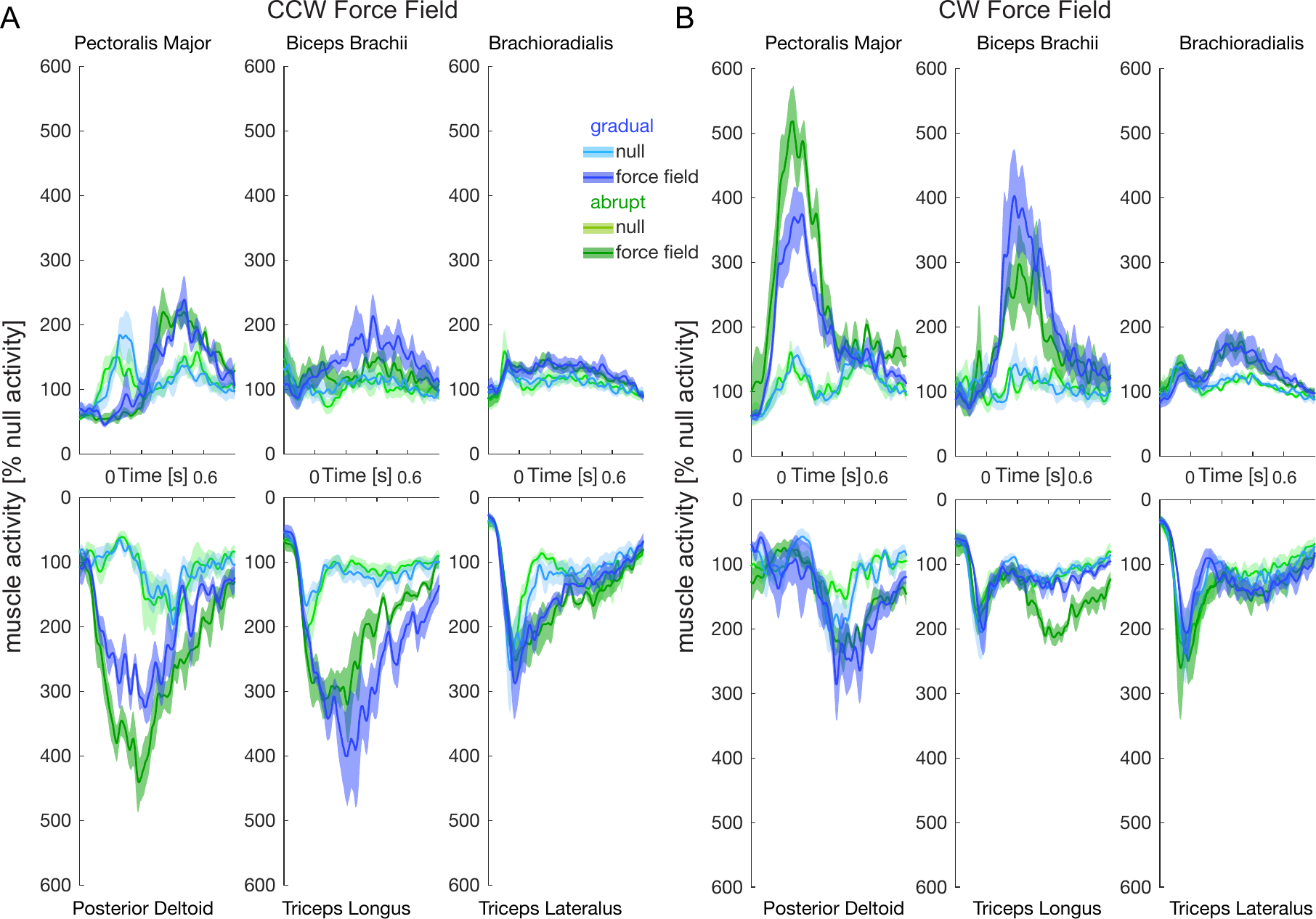}
\caption{Temporal profiles of muscle activity in the null force field and after adaptation for the abrupt (green) and gradual (blue) conditions. Data is only from the mechanical channel trials where the curl force field was not applied. Increasing extensor muscle activity is plotted in the downward direction. Prior to averaging across participants, the EMG has been smoothed with a 50-point (25ms) smoothing function. A: EMG profiles in the CCW curl force field. Null field activity (all 10 blocks in the pre-exposure phase) prior to adaptation is indicated by the light green and light blue traces. Final adaptation activity (all 20 blocks in the final exposure phase) is indicated by the dark green and dark blue traces. Muscle activity has been aligned to the start of the movement (0 s). Solid lines indicate mean across participants and shaded regions indicate s.e.m. B: EMG profiles in the CW curl force field.}
\end{figure}

The possibility that abrupt and gradual adaptation results in different patterns of adaptation is very intriguing but difficult to test here without a larger number of participants, as the participants adapted to both CW and CCW force fields. Nevertheless, we can perform some exploratory analysis by combining the CW and CCW force fields together. To do so, we combined the appropriate muscles (considering the agonist or antagonist role of the muscle in each force field). For example, the pectoralis major activity in the CW force field and posterior deltoid in the CCW force field were combined, and a t-test was used to contrast the mean level (over 100-500ms) in the abrupt and gradual conditions. We found a significantly larger muscle activity for the agonist shoulder muscle in the abrupt condition ($t_{22}=3.59, p=0.0016$) and a significantly larger muscle activity for the agonist biarticular muscle in the gradual condition ($t_{22}=2.27, p=0.0328$), but all other comparisons were not significantly different (all $p>0.5$). However, it is important to clarify that these statistics are exploratory and provide only weak support for this difference in final muscle activation patterns.

\begin{figure}
\centering
\includegraphics[width=12cm]{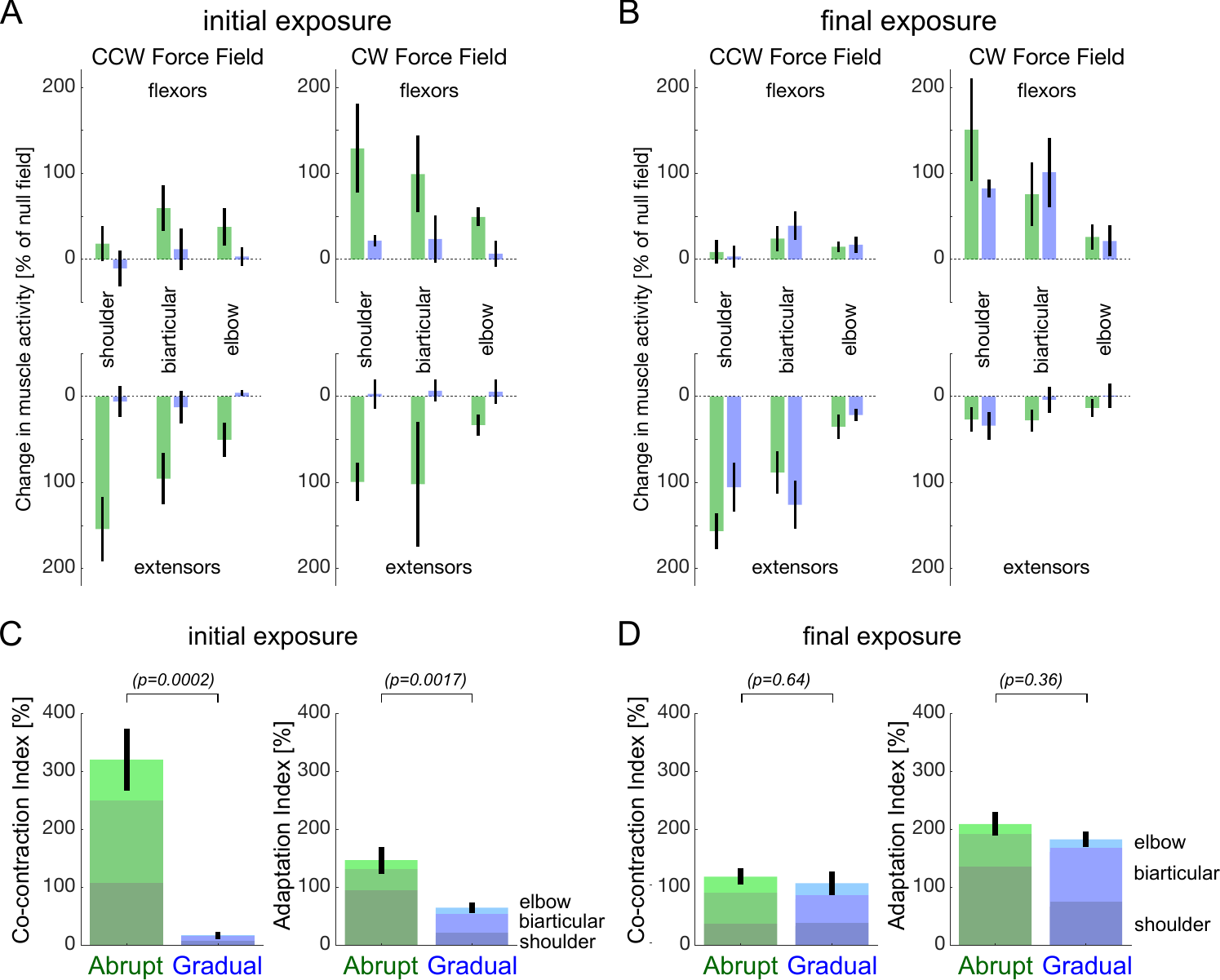}
\caption{Muscle activity in initial and final exposure to abrupt (green) and gradual (blue) application of force fields. Data is only from the mechanical channel trials where the curl force field was not applied. A: Initial exposure (first 10 blocks in initial exposure phase) to abrupt force field elicits large co-contraction in shoulder, biarticular and elbow muscles for both CCW and CW curl fields. Muscle activity is calculated as increase relative to the null field activity. Bars indicate mean ($\pm$ 95\% confidence intervals) integrated muscle activity from -100 to 600 ms from the start of the movement. If the error bars do not overlap then this indicates a significant difference at $p<0.05$. B: Final exposure (last 10 blocks in final exposure phase) shows similar levels of change in muscle activity for both abrupt and gradual change in dynamics. C: The co-contraction index and adaptation index of muscle activity in the initial exposure. Bar indicates total values across the muscle pairs ($\pm$ s.e.m.) and the colors indicate the relative contribution from the shoulder muscles (dark colors), biarticular muscles (medium colors) and elbow muscles (light colors). Values are across both the CCW and CW curl fields. Statistics indicate result of t-test. D: The co-contraction index and adaptation index of muscle activity in the final exposure periods.}
\end{figure}

Similar effects can be seen when we quantify the increases in EMG for both initial and final exposure (Fig. 5). As expected, the abrupt condition produced large increases in muscle activation for all six muscles (and therefore increased co-contraction) during the initial exposure (Fig. 5A). In contrast, only a small increase in muscle activity was seen initially in the gradual condition. However, during the final exposure phase we see a similar level of muscle activity in both abrupt and gradual conditions (Fig. 5B). Here again the difference between the abrupt and gradual adaptation appears in our experiment. Abrupt presentation of the force field appears to recruit higher activation in the shoulder muscles whereas gradual adaptation recruited higher levels of biarticular muscle activity. As error bars reflect the 95\% confidence intervals, some of these differences are significant. In order to quantify the degree of co-contraction and adaptation in the experiments across the gradual and abrupt conditions, we calculated the co-contraction and adaptation indices (Fig. 5C and 5D). The co-contraction index is a simple measure to capture the relative amount of activation in antagonistic muscle pairs, whereas the adaptation index is designed to indicate the amount of muscle activity in a specific direction (reciprocal activation) that might be directed to compensate for the force field. In the initial exposure phase (Fig. 5C), the co-contraction index was much higher in the abrupt condition than in the gradual condition ($t_{11}=5.3314, p=0.0002 ; BF_{10}=135.15 $). The adaptation index was also higher in the abrupt condition ($t_{11}=4.1241, p=0.0017 ; BF_{10}=25.75 $), but this difference was much smaller (approximately twice the level). However, in the final exposure phase (Fig. 5D), there were no significant differences in either the co-contraction ($t_{11}=0.4797, p=0.6409 ; BF_{10}=0.317 $) or adaptation ($t_{11}=0.9499, p=0.3625 ; BF_{10}=0.420 $) measures. The difference in the shoulder and biarticular muscle activity at the end of adaptation can again be seen in the adaptation index measure (Fig. 5D, right), but does not show up in the co-contraction index. Despite these differences, the overall muscle activity levels were similar across both presentations of the dynamics.

\subsection{Visuomotor Feedback Responses}

\begin{figure}
\centering
\includegraphics[width=11cm]{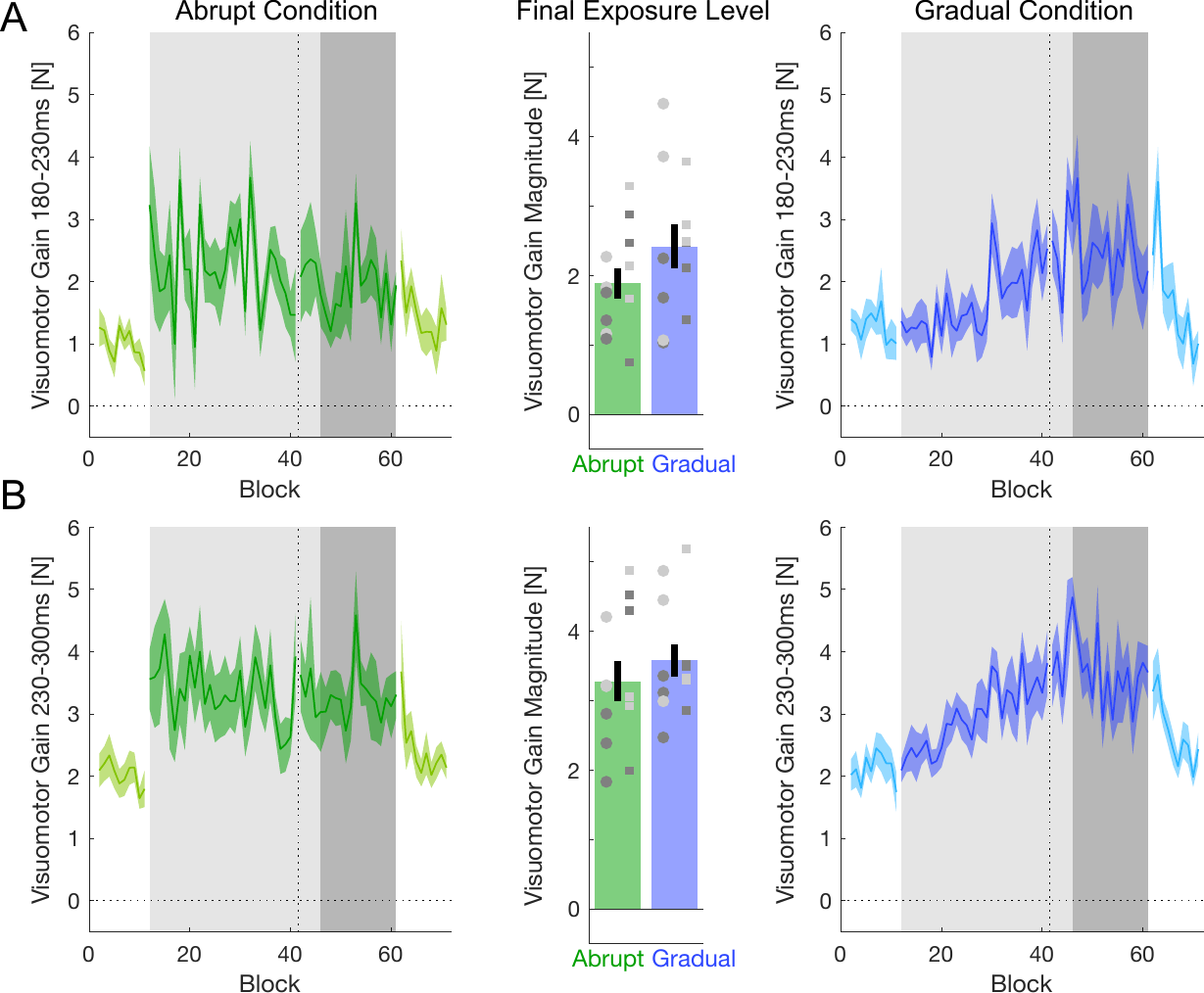}
\caption{Changes in visuomotor feedback gain during adaptation to abrupt (green) and gradual (blue) curl force fields. A: Visuomotor feedback gains during 180-230ms after the visual perturbation onset. Figure plotted as in Fig. 2. B: Visuomotor feedback gains during 230-300ms after the visual perturbation onset. Visuomotor feedback gains are measured on channel trials where the curl force fields were not applied.}
\end{figure}

Throughout the abrupt and gradual experiments, the visuomotor feedback responses were measured using probe trials in which the hand was constrained to a mechanical channel, but a visual perturbation of the cursor position was applied. The visuomotor feedback response was quantified over two intervals: an early interval between 180 and 230 ms (Fig. 6A) and a later interval between 230 and 300 ms (Fig. 6B). In both intervals the onset of the abrupt change in dynamics produced a rapid increase in the visuomotor gain (green traces) which then remained fairly high over the rest of the exposure phase, with a possible slight decrease over learning as seen in previous work \cite{franklin_visuomotor_2012}. In contrast, when the curl force field was applied gradually, there was little to no initial increase in the visuomotor gains, which instead gradually increased over the whole exposure period and then plateaued as the full level of force field was applied in the final exposure trials. Despite the very different patterns of visuomotor gains during the learning phase, the visuomotor gains in the last 15 blocks of the exposure phase were not significantly different between the abrupt and gradual conditions in either the early ($F_{1,10}=1.697, p=0.122; BF_{10}=0.922$; Fig. 6A bar plot) or late intervals ($F_{1,10}=0.542, p=0.334; BF_{10}=0.537$; Fig. 6B bar plot). For both cases, there was also no effect of condition order (early: $F_{1,10}=0.409, p=0.576; BF_{10}=0.665$; late: $F_{1,10}=0.531, p=0.519; BF_{10}=0.554$). Therefore, both conditions produced the same final level of visuomotor gains regardless of the large difference in kinematic errors during the adaptation. When the force field was removed abruptly in both conditions, we see an initially high visuomotor gain that decreased rapidly in these null field trials.

\begin{figure}
\centering
\includegraphics[width=9cm]{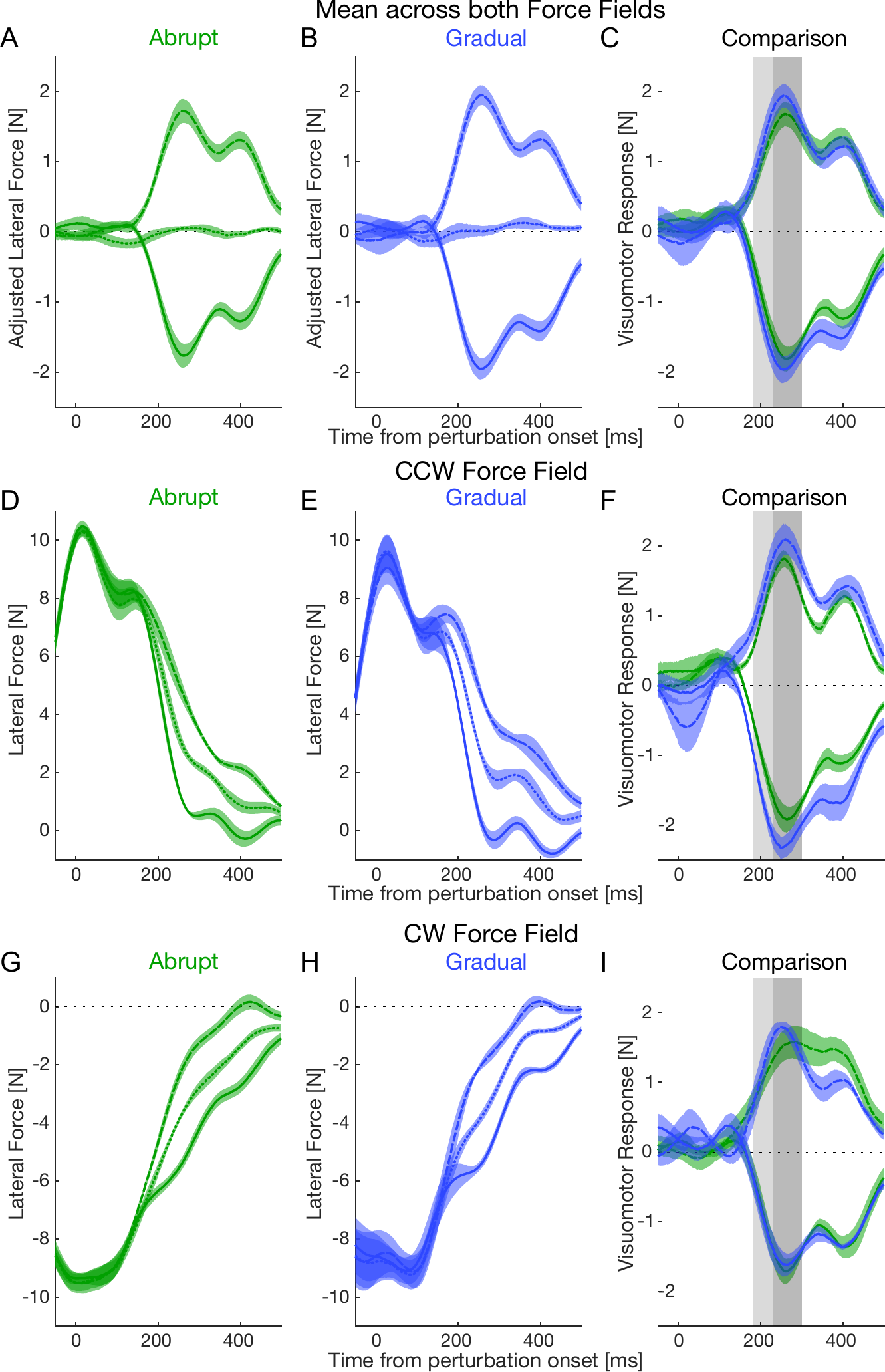}
\caption{Comparison of final visuomotor feedback responses after adaptation to the abrupt (green) and gradual (blue) curl force fields. A: Mean ($\pm$ s.e.m.) lateral force produced after rightward (solid lines), zero (dotted lines) and leftward (dashed lines) visual perturbations across both CCW and CW fields in the abrupt condition. Lateral force was adjusted by subtracting the mean lateral force across all channel trials in each force field. B: Lateral force in response to visual perturbations across both CCW and CW fields in the gradual condition. C: Visuomotor responses (zero perturbation subtracted) across both CCW and CW fields. Light grey shaded region indicates the early visuomotor response interval (180-230 ms) while the dark grey region indicates the late visuomotor response interval (230-300 ms). D: Lateral force produced in response to visual perturbations after adaptation to the abrupt onset of the CCW force field. E: Lateral force produced in response to visual perturbations after adaptation to the gradual onset of the CCW force field. F: Visuomotor responses in the CCW force field after abrupt and gradual adaptation. G-I: Visuomotor force responses after adaptation to the CW force field.}
\end{figure}

As previous studies have shown that movement kinematics can influence feedback gains \cite{crevecoeur_feedback_2013, cesonis_time--target_2020}, we also examined the peak velocities. The mean peak forward velocities ($\pm$ std) were 74.08 $\pm$ 10.15 m/s in the gradual condition and 75.45 $\pm$ 10.71 m/s in the abrupt condition. To test for differences across the adaptation period (early and late), we performed a repeated ANOVA. There was no difference between abrupt and gradual ($F_{1,11}=0.495, p=0.496$), or between early and late exposure phases ($F_{1,11}=1.919, p=0.193$), but there was a significant interaction effect ($F_{1,11}=6.042, p=0.032$). However post-hoc analysis found no differences between any of the conditions (all $p>0.098$).

In order to contrast the visuomotor feedback gains at the end of the exposure period we plotted the lateral hand force as a function of the time from perturbation onset (Fig. 7) in these probe trials. The lateral force is the force produced by the participant against the wall of the virtual channel. The force produced after the abrupt introduction of the force field (Fig. 7A) looks similar to that after the gradual introduction of the force field (Fig. 7B). When we subtract the zero-perturbation condition, we can see that the visuomotor response is similar in both conditions, not only in the early and late intervals but across the whole response (Fig. 7C). Participants in these experiments adapted to both the CCW and CW force fields, so we also examined the force response in each of these force fields separately (Fig. 7D-I). We find similar lateral forces against the channel wall for both the abrupt and gradual conditions in the CCW force field, but when we subtract the zero-perturbation condition it looks as though the forces are slightly larger in the gradual condition (Fig. 7F). In the CW force field, the lateral forces are in the opposite directions, but here the comparison shows a similar response in both abrupt and gradual conditions (Fig. 7I). One interesting point is that the CCW and CW force fields require opposite adaptive forces which can be clearly seen in the force traces (e.g. compare Fig. 7 D and G). In each force field a perturbation in one direction would be resisted by highly active muscles whereas in the other direction these muscles would have much lower activation (e.g. Fig. 5). This would then be reversed in the opposite force field. However, despite these differences the visuomotor force response is roughly equal in both perturbation directions. To support this claim, we compared the feedback force response to perturbations in the direction of the force field to those opposite to the direction of the force field using a t-test. We found no significant differences in either the early ($t_{22}=0.929, p=0.3629$) or late ($t_{22}=0.786, p=0.440$) intervals. This result supports our previous results that visuomotor feedback responses do not exhibit gain scaling \cite{franklin_visuomotor_2012, franklin_rapid_2017}, at least at the level of force responses.

\begin{figure}
\centering
\includegraphics[width=13cm]{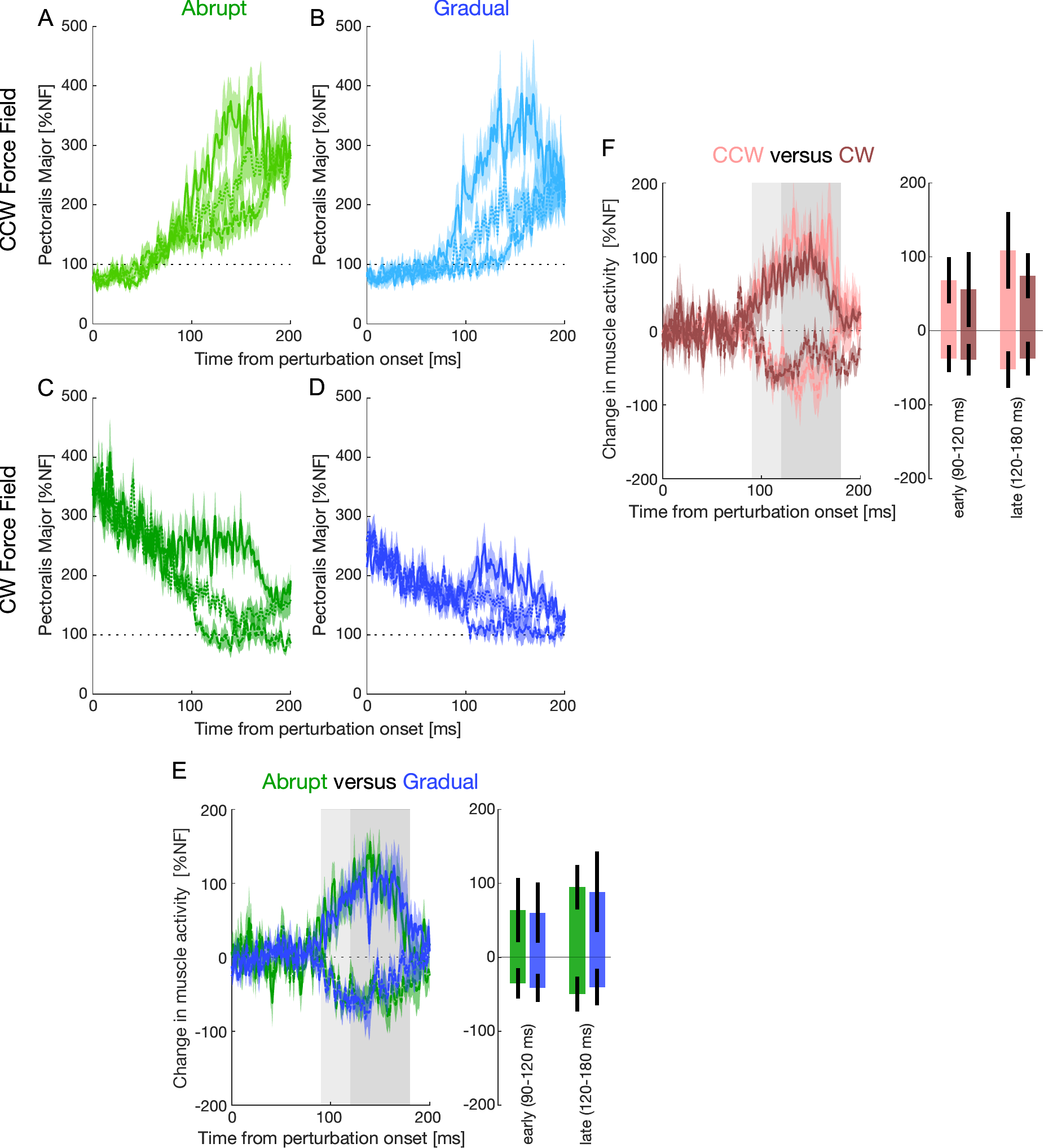}
\caption{Visuomotor feedback responses in the pectoralis major muscle after force field adaptation. A: Pectoralis major activity to leftward (dashed lines), zero (dotted lines), and rightward (solid lines) visual perturbations after abrupt adaptation to the CCW force field. Activity is scaled according to the level of muscle activity in the null field (mean between -50 and +50 ms prior to the perturbation time) which is represented by the dotted black line. Shaded region indicates the s.e.m. B: Muscle activity after gradual adaptation to the CCW force field. C: Muscle activity after abrupt adaptation to the CW force field. D: Muscle activity after gradual adaptation to the CW force field. E: Visuomotor responses (perturbation – zero perturbation) averaged across the CCW and CW force fields for the abrupt (green) and gradual (blue) conditions. Rightward perturbations (solid lines) produce an excitatory response whereas leftward perturbations (dashed lines) inhibit the muscle activity. Light grey and dark grey bars indicate the early (90-120 ms after perturbation onset) and late (120-180 ms) visuomotor response time windows. Bar plot quantify the responses over the early and late windows. Error bars indicate 95\% confidence intervals. F: Visuomotor responses averaged across abrupt and gradual conditions to examine differences between the CCW (pink) and CW (brown) force fields}
\end{figure}

\begin{figure}
\centering
\includegraphics[width=13cm]{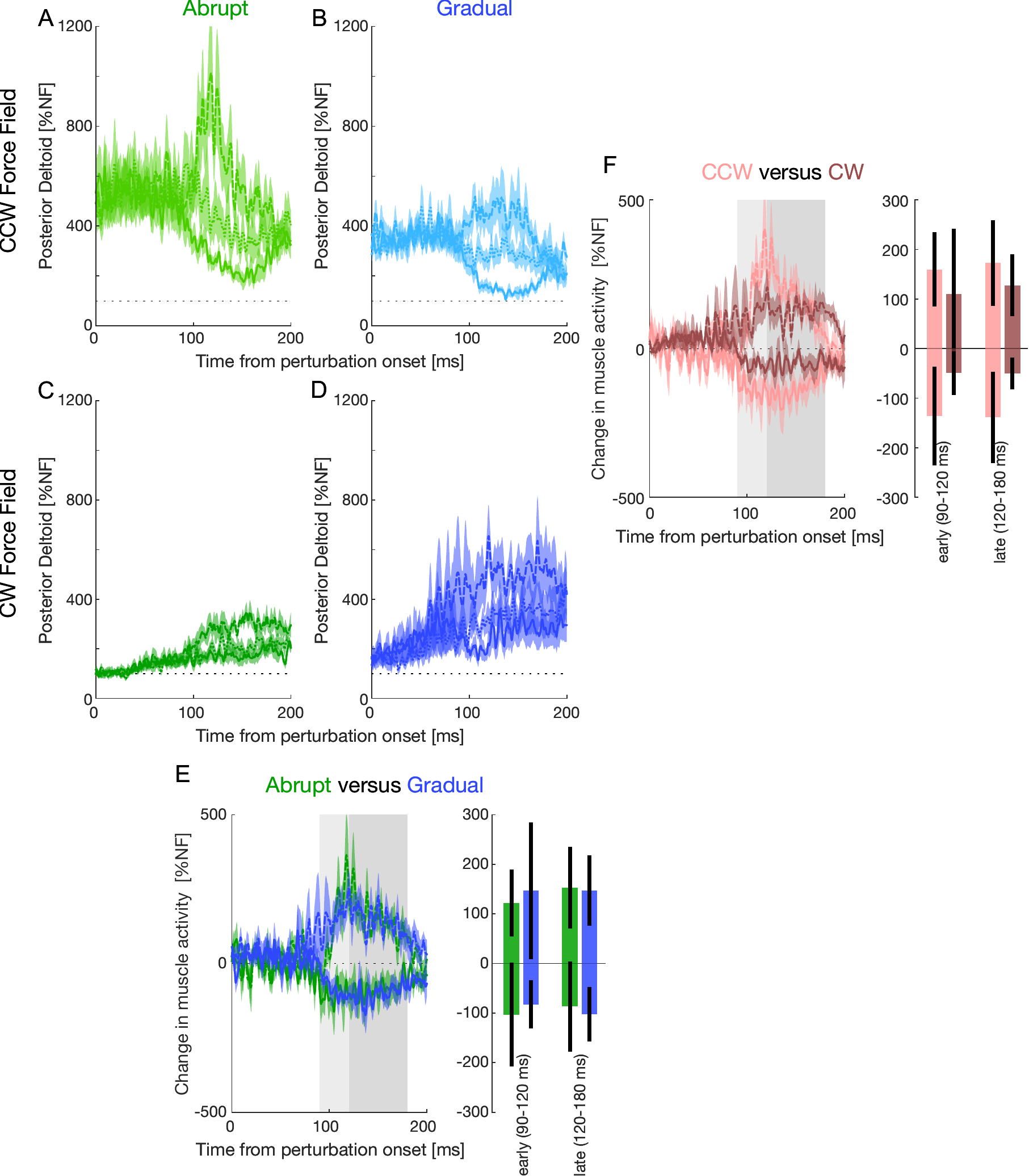}
\caption{Visuomotor feedback responses in the posterior deltoid muscle after force field adaptation. Responses plotted as in Fig. 8 to leftward (dashed lines), zero (dotted lines), and rightward (solid lines) visual perturbations. A: Posterior deltoid activity to leftward, zero, and rightward visual perturbations after abrupt adaptation to the CCW force field. B: Gradual adaptation to the CCW force field. C: Abrupt adaptation to the CW force field. D: Gradual adaptation to the CW force field. E: Visuomotor responses (perturbation – zero perturbation) averaged across the CCW and CW force fields for the abrupt (green) and gradual (blue) conditions. Bar plot shows responses over the early and late windows, with error bars indicating 95\% confidence intervals. F: Visuomotor responses averaged across abrupt and gradual conditions to examine differences between the CCW (pink) and CW (brown) force fields.}
\end{figure}

Finally, we examined the muscle responses to the visual perturbation, particularly those in the pectoralis major (Fig. 8) and posterior deltoid (Fig. 9); the major muscles to correct lateral perturbations in this movement. As the background load and muscle activity are different across force fields, the muscle responses to the visual perturbation is shown separately for the CCW and CW fields for both the abrupt and gradual conditions (Fig. 8A-D). Visual perturbations produce clear excitation or inhibition of the muscular activity. If we average across the two force fields we see that there are similar responses in the pectoralis major in both the abrupt and gradual conditions (Fig. 8E), with no differences across either the early or late visuomotor response windows (error bars represent 95\% confidence intervals). If instead, we average across the abrupt and gradual conditions, we can directly compare the muscular responses in the CCW versus the CW force fields (Fig. 8E). Here again there were no differences across the temporal response or within the early or late visuomotor response windows. That is, despite differences in the background loads of the muscles, the visuomotor response at the end of the learning the force fields were similar for both CCW and CW force fields.  Similar responses were observed in the posterior deltoid (Fig. 9). When averaged across the force fields, we found no differences in the muscular responses between the abrupt and gradual conditions (Fig. 9E). However, when averaging across abrupt and gradual conditions, we found apparent differences in the muscular responses between the CCW and CW force fields, although these differences were not statistically significant (Fig. 9F). Visual feedback responses may have exhibited small field-specific differences across the abrupt and gradual fields (for example compare posterior deltoid response in Figs. 9A and 9B). However, with only six participants available for these specific comparisons, it is not possible to examine these possible trends further in this study.

\section{Discussion}
The goal of this study was to examine whether internal model uncertainty drives changes in the feedback gains during adaptation. In order to do this, participants performed reaching movements where the environmental dynamics were either changed abruptly (producing large kinematic errors) or gradually (producing small kinematic errors). Abrupt changes in dynamics produced large kinematic errors during the movements signaling that the internal model was incorrect – which should increase internal model uncertainty. In contrast, a gradual change in the dynamics produced only small kinematic errors and should produce much lower uncertainty in the internal model. In the initial exposure to the force fields, participants in the abrupt condition experienced large kinematic errors, extensive muscle co-contraction, increased visuomotor feedback gains, and a rapid increase in force compensation. In contrast, participants in the gradual condition experienced little to no kinematic errors, very little co-contraction and only small increases in the visuomotor feedback gains. Despite this, the force compensation was maintained at around 70\% of the applied force field throughout the initial exposure period, and was associated with the respective changes in the muscle adaptation index. Although there were large differences in the initial exposure phase, participants in both abrupt and gradual conditions reached similar levels of kinematic error, force compensation, co-contraction, muscle activity and visuomotor feedback gains by the end of the exposure phase. 

Here we claim that internal model uncertainty drives the changes in feedback gains, rather than internal model inaccuracy. It is important to point out that we do not have a specific measure of internal model uncertainty, and that here we only manipulate it with increased errors through changes in the environmental dynamics which also directly affects internal model accuracy. Despite this, we believe that it is internal model uncertainty rather than inaccuracy that drives these changes. First of all, we may have internal model inaccuracy, but without errors indicating that this is the case, we predict no increased feedback gains. For example, we can consider the first trial in the exposure phase. There is initially high internal model inaccuracy as the prediction is a null force field (but environment is a force field), but low internal model uncertainty as the current model has accurately predicted all the previous trials. Only after a large error has indicated that the internal model is inaccurate, directly increasing internal model uncertainty, will we find increased feedback gains. Moreover, we predict higher feedback gains if the internal model uncertainty is high even if the internal model accuracy is perfect on a given trial (for example as the external dynamics change on a trial-by-trial basis). While we do not yet know any data supporting this testable prediction, the effects are similar to the changes in grip force seen during adaptation to changing force fields \cite{hadjiosif_flexible_2015}. In this paper \cite{hadjiosif_flexible_2015}, Hadjiosif and Smith showed that the grip force was much more sensitive to increased variability in the strength of the force fields than in the size of the force field, suggesting a connection between statistical confidence and the grip force matching similar changes in co-contraction. Here we also find similar changes in feedback gains which match their predictions. Our interpretation that internal model uncertainty drives increases in feedback gains also predicts that we would find a brief increase in feedback gains when the force field was removed -- initial washout trials -- as we found in our previous work \cite{franklin_visuomotor_2012}. However, there is no clear evidence of this effect in the current manuscript, where the initial feedback gains in the post-exposure phase are similar to those in the exposure phase (Fig. 6).

In this study we used a within-subject design, with a counterbalanced order of presentation such that half of the participants experience the gradual adaptation first and half the abrupt adaptation. A limitation of this design is that participants have already experienced adapting to a force field upon presentation of the second force field. This effect was limited by ensuring that participants adapted to the opposite force field during the second adaptation period (avoiding the effect of savings), and by having 303 null field trials between the first adaptation and the second adaptation phases to ensure sufficient washout. Moreover, there were no statistically significant effects of experiment order. However, there are also important advantages to this design. First of all, the amount to which each participant adapts to a force field varies greatly, with some participants adapting almost fully and others only partially \cite{wu_temporal_2014, Cluff_tradeoffs_2019}. The repeated measures design allows us to adjust for participant variability in the adaptation process. More importantly this design allows us to keep the EMG electrodes on the participant to record muscle activity in both gradual and abrupt conditions. Overall, this design improves our sensitivity to any differences in the adaptation process without requiring a large number of participants. Even though we only have 12 participants, limiting the effect of individual variance by having all participants perform all conditions improves estimates of any effects. However, counterbalancing the force fields means that most of the muscle activity data is presented for six participants (CW and CCW separately), and thereby limits our ability to statistically test these results further. Here we have analyzed and presented EMG data normalized to the activity in the null field as in our previous work \cite{franklin_adaptation_2003, franklin_visuomotor_2012}. While this normalization technique allows for clear comparison across different papers and experiments, it has some limitations in the normalization between muscles. That is if one muscle has very low levels of activation throughout the whole movement, this muscle is more susceptible to increased noise. As with any summed measure of muscle activity across muscles, care needs to be taken with interpretations of our indexes of co-contraction and adaptation which sum activity across muscles. However, these value along with the trial-by-trial and temporal patterns of muscle activity for each muscle individually provide evidence that there are similar levels of muscle activity after adaptation to both gradual and abrupt changes in dynamics.

Many studies have compared gradual adaptation with abrupt adaptation to force fields \cite{malfait_is_2004, klassen_learning_2005, kluzik_reach_2008, huang_persistence_2009, orban_de_xivry_contributions_2011, pekny_protection_2011, milner_different_2018, alhussein_dissociating_2019}. It has been shown that the final level of adaptation is similar regardless of whether the novel dynamics are presented abruptly or gradually \cite{malfait_is_2004, klassen_learning_2005, milner_different_2018, alhussein_dissociating_2019}. Here we also found no difference in the final level of kinematic error (MPE) or force compensation, agreeing with these previous studies. It has been suggested that gradual presentation of novel dynamics drives changes in the internal model of the limbs dynamics rather than in the internal model of the tool (or robot)  \cite{kluzik_reach_2008}, which could explain why the motor memory formed with gradual adaptation does not transfer bimanually to the other limb \cite{malfait_is_2004}. Although it has been suggested that gradual presentation produces better retention \cite{huang_persistence_2009}, more recent studies have shown that the retention rates are similar, and that the biggest effect on retention is simply the amount of training \cite{alhussein_dissociating_2019}. Nevertheless, as larger errors are more likely to induce adaptation of a new motor memory \cite{oh_minimizing_2019} or adapt a specific tool related motor memory \cite{kluzik_reach_2008}, we might expect that there are subtle differences in the properties of the motor memories that are formed under these two different situations, such as the fact that gradually adapted motor memories do not transfer across limbs \cite{malfait_is_2004}.

Although earlier studies suggested that abrupt presentation of novel dynamics is learned faster \cite{huang_persistence_2009}, more recently it was claimed that the learning rate in abrupt and gradual dynamics is similar \cite{milner_different_2018}. Milner and colleagues found that while amplitude error decreased faster under abrupt conditions, the temporal error or smoothness decreased faster under gradual conditions. Here we did not compare the rate of adaptation under the two conditions, however the force compensation calculated relative to the presented dynamics was consistently 70\% of the applied force field throughout both the gradual increase of the force field and the steady state phase. Under the abrupt change in dynamics, the force compensation increased up to a similar level. Therefore, it appears that under both conditions, the adaptation mechanism is able to incorporate a similar level of the error to modifying the predictive force compensation. Previous work has shown that the adaptation system is more sensitive to small errors than larger errors \cite{fine_motor_2006, wei_relevance_2009, marko_sensitivity_2012, hayashi_divisively_2020}, which might suggest that gradual dynamics would be learned faster. Here we have shown that the large errors produced during abrupt onset of dynamics induces much higher levels of co-contraction, which have been suggested to increase the speed of adaptation \cite{heald_increasing_2018}. It has been suggested that co-contraction might increase the rate of adaptation by concentrating the adaptation within the range of state space in which the dynamics must finally be learned \cite{heald_increasing_2018}, as participants learn adaptation of dynamics as a function of the visited states rather than planned states \cite{gonzalez_castro_binding_2011}. However, the small kinematic errors that occur during gradual presentation of the force field also mean that participants are never perturbed away from the region of state space to be learned. Thus, gradual presentation could be learned just as quickly as abrupt presentation despite the absence of co-contraction. The similar rates of adaptation under these different conditions likely arises through trade-offs in the competing mechanisms of co-contraction, error sensitivity, and nearness to the learned state space.

Abrupt adaptation to novel dynamics has been shown to cause a large increase in initial co-contraction which gradually reduces as the internal model is gradually learned \cite{thoroughman_electromyographic_1999, franklin_adaptation_2003, milner_impedance_2005, huang_reduction_2012}. This increase in co-contraction from errors is so strong, that it has been shown as a response to an abrupt introduction of a visuomotor rotation \cite{huang_reductions_2014}, despite the fact that co-contraction cannot reduce this type of visual transformation errors. Here we also showed a large initial increase in co-contraction when the dynamics were abruptly applied, but little or no increase in co-contraction when the dynamics were only gradually applied, even though both conditions show adaptation to the force field in terms of force compensation and our EMG based adaptation index. This suggests that it is the errors that drive these changes in co-contraction and not the adaptation process. Despite these differences early in learning, we found similar levels of co-contraction at the end of the exposure phase in both conditions. Therefore, even when co-contraction is not induced through large kinematic errors, co-contraction gradually builds up as part of the adaptation process.

The final levels of muscle co-contraction were similar after both gradual and abrupt adaptation to the dynamics, and clearly different from the levels of co-contraction in the null force field as shown previously \cite{darainy_muscle_2008, franklin_visuomotor_2012}. However, the actual pattern of muscle activation after adaptation may have varied depending on the manner in which the dynamics were learned. However, here we only have weak statistical support for these differences, so further experimentation will be required to determine whether this difference is consistent. In the abrupt condition participants tended to have higher shoulder muscle activity, whereas in the gradual condition participants tended to have higher biarticular muscle activity. We suggest two possible reasons for these differences in the final pattern of muscle activation. First, it is possible that abrupt onset of novel dynamics produces large errors, but these errors can be clearly associated with a specific cause – the robot is producing a disturbance which is perturbing specific muscles more than others (in this case single joint shoulder muscles). The sensorimotor control system therefore associates the errors with this novel task, forming motor memories of the tool rather than the body \cite{berniker_estimating_2008, kluzik_reach_2008}. On the other hand, in the gradual condition there are small continuous errors which reduce the task success \cite{pekny_protection_2011} but cannot be associated with any specific cause, inducing participants to adapt their baseline or limb dynamics model \cite{kluzik_reach_2008, oh_minimizing_2019}. These small errors and reduced task success may be signals of instability \cite{crevecoeur_movement_2010, burdet_stability_2006} that drive specific increases in biarticular muscle activation as these have specific critical roles in limb stability \cite{mcintyre_control_1996, franklin_adaptive_2003}. As there is extensive redundancy in the arm muscles that could be used to compensate for the external loads of the force fields, different conditions would drive different final adaptation measures. The second possible explanation for these findings is that adaptation occurs through a feedback error learning mechanism \cite{kawato_hierarchical_1987, franklin_cns_2008, albert_neural_2016}. In this case, different errors produce different sensory feedback signals which are then incorporated into the feedforward motor command on the subsequent trial. A different pattern of feedback responses would therefore result in a different final pattern of muscle activity. Such a possibility could be connected to potential different feedback responses in the different environments that we were not able to examine with the current study as discussed with Fig 9. That is, the first possibility suggests that large errors directly select different models whereas consistent small errors may signal instability leading to differential adaptation. In contrast, the second possibility suggests that it is simply the different pattern of errors that drives these differences. These two possibilities are not mutually exclusive, so the different patterns of muscle activity could be driven by both. However, further experiments are required to determine whether these results are a general result of gradual versus abrupt adaptation or if they are specific to the movement direction studied in this experiment.  A consistent finding of higher biarticular muscle activation would be important for use in rehabilitation studies \cite{patton_robot-assisted_2004, huang_augmented_2012, reinkensmeyer_computational_2016} where the goal is training new patterns of muscle activity. Gradual adaptation in a split-belt treadmill has already shown to generalize better to normal walking \cite{torres-oviedo_natural_2012}, and an associated increase in stabilizing biarticular muscle activation could be an additional helpful result of such training if this is a consistent finding.

Throughout the experiment, we assessed the gain of the visuomotor feedback response using rapid perturbations of the visual representation of the hand \cite{brenner_fast_2003, sarlegna_target_2003, saunders_humans_2003}. These visuomotor feedback gains demonstrate task modulation \cite{knill_flexible_2011} and are tuned to changes in the dynamics of the environment \cite{franklin_rapid_2017}. Our previous work showed that early in learning novel dynamics (or when these dynamics are suddenly removed) the visuomotor feedback gains are upregulated \cite{franklin_visuomotor_2012} and then gradually reduce to a plateau after adaptation is completed. In the curl force field, the final level of visuomotor feedback gain was higher than in the null field, which was interpreted as adaptation to the increased uncertainty in the curl force field. We hypothesized that this initial increase in feedback gains was driven by increased uncertainty in the internal model of the dynamics \cite{franklin_visuomotor_2012}, causing large feedback gains similar to the increased co-contraction seen in early adaptation \cite{thoroughman_electromyographic_1999, franklin_adaptation_2003}. Here we tested this theory by having participants adapt to abrupt and gradual force field presentations. As in our previous work, the abrupt onset or removal of the force field elicited large increases in feedback gains. However, when the force field was only applied gradually these feedback gains remained low and only increased slowly to the same plateau level after adaptation was complete. However, both the force field compensation and adaptation index show that adaptation to the force field already occurs early in gradual learning, demonstrating that this initial upregulation of feedback gains is not just a side effect of adaptation. Instead, we suggest that it represents a reactive increase in feedback gains due to the uncertainty in the internal model as signaled by large kinematic errors.

Feedback gains during adaptation have also been studied using stretch reflexes \cite{cluff_rapid_2013, coltman_time_2020}, although examining the changes in stretch reflex gains independent of their gain scaling \cite{pruszynski_temporal_2009} is more difficult. By examining changes in the long latency stretch reflex response for a movement direction in between two movements in which force fields were learned, \cite{cluff_rapid_2013} showed that the long latency stretch reflex increased slowly as learning occurred, paralleling the slow increase in the visuomotor feedback gain that we found during gradual adaptation. On the other hand, a recent study \cite{coltman_time_2020} found only initial increases in the feedback gain during early learning, but no long-term changes after complete adaptation. However, the perturbation to assess the feedback responses was performed in the dwell period prior to the movement onset, so no long-term upregulation of the feedback gain during this time period would be needed for adaptation. Although this early increase in the feedback gains was attributed to the fast process of a two-state model of motor adaptation \cite{coltman_time_2020}, we believe that the rapid increase in feedback gains, particularly as seen here and previously \cite{franklin_visuomotor_2012}, could be better described as a reactive increase in response to error rather than as part of the adaptation process itself. This is particularly likely for feedback gains seen prior to movement initiation when little or no adaptation is ever found in the predictive force against the channel wall \cite{joiner_linear_2011, alhussein_dissociating_2019}. Further supporting our interpretation, we find identical levels of visuomotor feedback gain to perturbations to the left or right (see Fig. 7), even though adaptation as predicted by the fast process shows a directionality according to the force field.

\begin{figure}
\centering
\includegraphics[width=8cm]{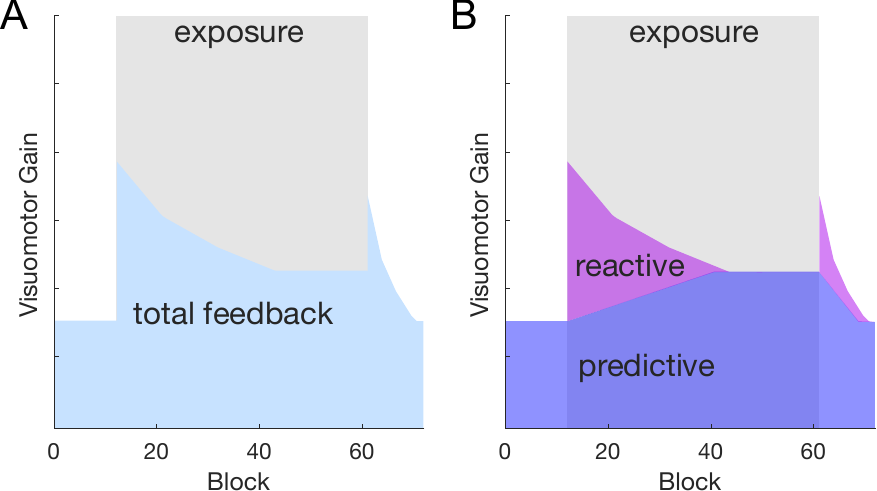}
\caption{Schematic of feedback changes during adaptation to novel dynamics. A: The pattern of feedback gain modulation during adaptation and de-adaptation to changes in environmental dynamics. Increases in feedback gain occur when the dynamics are changed (onset or offset), and the final level of feedback gain after adaptation to the force field is different to that in the null force field. B: We propose that the total feedback gain is comprised of the reactive (purple) and predictive (blue) changes in feedback gains during adaptation. The reactive feedback gains increase immediately in response to large errors (model predictive errors) and gradually reduce as learning occurs. The predictive feedback gains gradually increase (or decrease) during adaptation as they are tuned to the environment.}
\end{figure}

Overall, feedback gains during adaptation to novel dynamics demonstrate a clear pattern of modulation (Fig. 10A), with an initial rapid increase associated with large kinematic errors, followed by a gradual reduction to a new baseline level during the exposure. We suggest that this pattern is actually composed of two complementary processes (Fig. 10B) which we term reactive and predictive feedback gains \cite{franklin_rapid_2017}. Any large errors, signaling model uncertainty, produce rapid increases in reactive feedback gains, which are gradually reduced as learning occurs. We propose that after adaptation is complete, these contribute little or nothing to the overall feedback responses, where this pattern matches perfectly the changes found in a recent paper \cite{coltman_time_2020}. However, throughout the adaptation process, feedback responses are also learned and tuned according to the dynamics of the environment \cite{franklin_rapid_2017}, becoming predictive feedback gains which are part of the learned motor memory to compensate for the force field \cite{franklin_endpoint_2007, wagner_shared_2008, ahmadi-pajouh_preparing_2012, maeda_feedforward_2018}. The time course of these predictive change in the feedback gains matches well the changes seen in Cluff and Scott \cite{cluff_rapid_2013} where no reactive responses were required as no errors were experienced on these trials. These two contributions to the overall pattern of feedback control would explain the different results found in a variety of experiments \cite{franklin_visuomotor_2012, cluff_rapid_2013, coltman_time_2020}. Furthermore, we predict that each of these two components would have different properties; reactive feedback gains are likely to be broader in terms of temporal timing (before and after the movement) whereas predictive feedback gains may be more likely to generalize spatially to nearby movements similar to the generalization of predictive force \cite{shadmehr_adaptive_1994, malfait_transfer_2002, berniker_motor_2014}.

Recent work examining sensorimotor control have suggested robust control as a framework for explaining and interpreting human behaviour \cite{ueyama_mini-max_2014, crevecoeur_robust_2019, bian_model-free_2020}. Robust control would predict higher feedback gains with increased internal model uncertainties, fitting well our current experimental results. Moreover, it has been shown that each participant modulates feedback responses differently, leading to the suggestion that each participant may fall along a continuum between optimal and robust control \cite{Cluff_tradeoffs_2019}. It could be imagined that the reactive feedback gains might correspond to a robust control policy that attempts to provide stability despite internal model inaccuracies, and that the predictive feedback gains could reflect a shifting of the control policy to an energy efficient mode as the internal model accuracy is improved. However, it is unclear whether such shifting control policies are needed to explain the reactive and predictive feedback gains, and we do not examine variations in the feedback responses across participants. Nevertheless, the large up-regulation of feedback gains in response to large errors well matches the predictions of robust control \cite{Cluff_tradeoffs_2019, crevecoeur_robust_2019}.

Abrupt and gradual adaptation produce very different initial patterns of force compensation, muscle activity, co-contraction and feedback gains, but finally result in a similar pattern after adaptation despite the different time course of errors. Although the final adaptation is similar, there still remain subtle differences in terms of the pattern of muscle activity (Fig. 4\&5) and generalization \cite{malfait_is_2004}. However, the different patterns of visuomotor feedback regulation allow us to separate out two components of feedback regulation: reactive and predictive. Here we argue that internal model uncertainty drives upregulation of the reactive feedback gains, and that adaptation tunes the predictive feedback gains according to the environment.

\section*{ACKNOWLEDGEMENTS}
We thank Daniel Wolpert and James Ingram for support with the experimental facilities. 

\section*{CONFLICT OF INTEREST}
The authors declare no competing financial interests.

\section*{DATA AVAILABILITY}
Data and code for analysis is available at: https://doi.org/10.6084/m9.figshare.12816086.v1

\printendnotes

\bibliography{main}

\begin{thebibliography}{78}
\providecommand{\natexlab}[1]{#1}
\providecommand{\url}[1]{\texttt{#1}}
\providecommand{\urlprefix}{}

\bibitem[{Lackner and Dizio(1994)Lackner, J. R. and Dizio,
  P.}]{lackner_rapid_1994}
Lackner JR, Dizio P.
\newblock Rapid adaptation to {Coriolis} force perturbations of arm trajectory.
\newblock Journal of Neurophysiology 1994;72:299--313.

\bibitem[{Shadmehr and Mussa-Ivaldi(1994)Shadmehr, R. and Mussa-Ivaldi, F.
  A.}]{shadmehr_adaptive_1994}
Shadmehr R, Mussa-Ivaldi FA.
\newblock Adaptive representation of dynamics during learning of a motor task.
\newblock Journal of Neuroscience 1994;14:3208--3224.

\bibitem[{Conditt et~al.(1997)Conditt, M. A. and Gandolfo, F. and Mussa-Ivaldi,
  F. A.}]{conditt_motor_1997}
Conditt MA, Gandolfo F, Mussa-Ivaldi FA.
\newblock The motor system does not learn the dynamics of the arm by rote
  memorization of past experience.
\newblock Journal of Neurophysiology 1997;78:554--560.

\bibitem[{Thoroughman and Shadmehr(1999)Thoroughman, K. A. and Shadmehr,
  R.}]{thoroughman_electromyographic_1999}
Thoroughman KA, Shadmehr R.
\newblock Electromyographic correlates of learning an internal model of
  reaching movements.
\newblock The Journal of Neuroscience 1999;19:8573--8588.

\bibitem[{Osu et~al.(2002)Osu, Rieko and Franklin, David W. and Kato, Hiroko
  and Gomi, Hiroaki and Domen, Kazuhisa and Yoshioka, Toshinori and Kawato,
  Mitsuo}]{osu_short-_2002}
Osu R, Franklin DW, Kato H, Gomi H, Domen K, Yoshioka T, et~al.
\newblock Short- and long-term changes in joint co-contraction associated with
  motor learning as revealed from surface {EMG}.
\newblock Journal of Neurophysiology 2002;88(2):991--1004.

\bibitem[{Franklin et~al.(2003)Franklin, David W. and Osu, Rieko and Burdet,
  Etienne and Kawato, Mitsuo and Milner, Theodore
  E.}]{franklin_adaptation_2003}
Franklin DW, Osu R, Burdet E, Kawato M, Milner TE.
\newblock Adaptation to stable and unstable dynamics achieved by combined
  impedance control and inverse dynamics model.
\newblock Journal of Neurophysiology 2003;90(5):3270--3282.

\bibitem[{Milner and Franklin(2005)Milner, Theodore E. and Franklin, David
  W.}]{milner_impedance_2005}
Milner TE, Franklin DW.
\newblock Impedance control and internal model use during the initial stage of
  adaptation to novel dynamics in humans.
\newblock Journal of Physiology 2005;567(2):651--664.

\bibitem[{Hoffer and Andreassen(1981)Hoffer, J. A. and Andreassen,
  S.}]{hoffer_regulation_1981}
Hoffer JA, Andreassen S.
\newblock Regulation of soleus muscle stiffness in premammillary cats:
  intrinsic and reflex components.
\newblock Journal of Neurophysiology 1981;45(2):267--285.

\bibitem[{Kearney et~al.(1997)Kearney, R. E. and Stein, R. B. and Parameswaran,
  L.}]{kearney_identification_1997}
Kearney RE, Stein RB, Parameswaran L.
\newblock Identification of intrinsic and reflex contributions to human ankle
  stiffness dynamics.
\newblock IEEE Transactions on Biomedical Engineering 1997;44(6):493--504.

\bibitem[{Mirbagheri et~al.(2000)Mirbagheri, M. M. and Barbeau, H. and Kearney,
  R. E.}]{mirbagheri_intrinsic_2000}
Mirbagheri MM, Barbeau H, Kearney RE.
\newblock Intrinsic and reflex contributions to human ankle stiffness:
  variation with activation level and position.
\newblock Experimental Brain Research 2000;135(4):423--436.

\bibitem[{Akazawa et~al.(1983)Akazawa, K. and Milner, T. E. and Stein, R.
  B.}]{akazawa_modulation_1983}
Akazawa K, Milner TE, Stein RB.
\newblock Modulation of reflex {EMG} and stiffness in response to stretch of
  human finger muscle.
\newblock Journal of Neurophysiology 1983;49(1):16--27.

\bibitem[{Saliba et~al.(2020)Saliba, Christopher M. and Rainbow, Michael J. and
  Selbie, W. Scott and Deluzio, Kevin J. and Scott, Stephen
  H.}]{saliba_co-contraction_2020}
Saliba CM, Rainbow MJ, Selbie WS, Deluzio KJ, Scott SH.
\newblock Co-contraction uses dual control of agonist-antagonist muscles to
  improve motor performance.
\newblock bioRxiv 2020;p. 2020.03.16.993527.

\bibitem[{Franklin et~al.(2012)Franklin, Sae and Wolpert, Daniel M. and
  Franklin, David W.}]{franklin_visuomotor_2012}
Franklin S, Wolpert DM, Franklin DW.
\newblock Visuomotor feedback gains upregulate during the learning of novel
  dynamics.
\newblock Journal of Neurophysiology 2012;108(2):467--478.

\bibitem[{Coltman and Gribble(2020)Coltman, Susan K. and Gribble, Paul
  L.}]{coltman_time_2020}
Coltman SK, Gribble PL.
\newblock Time course of changes in the long-latency feedback response
  parallels the fast process of short-term motor adaptation.
\newblock Journal of Neurophysiology 2020;124(2):388--399.

\bibitem[{Oh and Schweighofer(2019)Oh, Youngmin and Schweighofer,
  Nicolas}]{oh_minimizing_2019}
Oh Y, Schweighofer N.
\newblock Minimizing {Precision}-{Weighted} {Sensory} {Prediction} {Errors} via
  {Memory} formation and switching in motor adaptation.
\newblock Journal of Neuroscience 2019;39(46):9237--9250.

\bibitem[{Franklin et~al.(2017)Franklin, Sae and Wolpert, Daniel M. and
  Franklin, David W.}]{franklin_rapid_2017}
Franklin S, Wolpert DM, Franklin DW.
\newblock Rapid visuomotor feedback gains are tuned to the task dynamics.
\newblock Journal of Neurophysiology 2017;118(5):2711--2726.

\bibitem[{Cluff and Scott(2013)Cluff, Tyler and Scott, Stephen
  H.}]{cluff_rapid_2013}
Cluff T, Scott SH.
\newblock Rapid feedback responses correlate with reach adaptation and
  properties of novel upper limb loads.
\newblock Journal of Neuroscience 2013;33(40):15903--15914.

\bibitem[{Malfait and Ostry(2004)Malfait, Nicole and Ostry, David
  J.}]{malfait_is_2004}
Malfait N, Ostry DJ.
\newblock Is interlimb transfer of force-field adaptation a cognitive response
  to the sudden introduction of load?
\newblock Journal of Neuroscience 2004;24(37):8084--8089.

\bibitem[{Klassen et~al.(2005)Klassen, Jessica J. and Tong, Christine C. and
  Flanagan, J. Randall JR}]{klassen_learning_2005}
Klassen JJ, Tong CC, Flanagan JRJ.
\newblock Learning and recall of incremental kinematic and dynamic sensorimotor
  transformations.
\newblock Experimental Brain Research 2005;164(2):250--259.

\bibitem[{Kluzik et~al.(2008)Kluzik, JoAnn and Diedrichsen, Jörn and Shadmehr,
  Reza and Bastian, Amy J.}]{kluzik_reach_2008}
Kluzik J, Diedrichsen J, Shadmehr R, Bastian AJ.
\newblock Reach adaptation: what determines whether we learn an internal model
  of the tool or adapt the model of our arm?
\newblock Journal of Neurophysiology 2008;100(3):1455--1464.

\bibitem[{Huang and Shadmehr(2009)Huang, V. S. and Shadmehr,
  R.}]{huang_persistence_2009}
Huang VS, Shadmehr R.
\newblock Persistence of motor memories reflects statistics of the learning
  event.
\newblock Journal of Neurophysiology 2009;102(2):931--940.

\bibitem[{Orban~de Xivry et~al.(2011)Orban de Xivry, Jean-Jacques and
  Criscimagna-Hemminger, Sarah E. and Shadmehr,
  Reza}]{orban_de_xivry_contributions_2011}
Orban~de Xivry JJ, Criscimagna-Hemminger SE, Shadmehr R.
\newblock Contributions of the motor cortex to adaptive control of reaching
  depend on the perturbation schedule.
\newblock Cerebral Cortex 2011;21(7):1475--1484.

\bibitem[{Pekny et~al.(2011)Pekny, Sarah E. and Criscimagna-Hemminger, Sarah E.
  and Shadmehr, Reza}]{pekny_protection_2011}
Pekny SE, Criscimagna-Hemminger SE, Shadmehr R.
\newblock Protection and expression of human motor memories.
\newblock Journal of Neuroscience 2011;31(39):13829--13839.

\bibitem[{Milner et~al.(2018)Milner, Theodore E. and Firouzimehr, Zeinab and
  Babadi, Saeed and Ostry, David J.}]{milner_different_2018}
Milner TE, Firouzimehr Z, Babadi S, Ostry DJ.
\newblock Different adaptation rates to abrupt and gradual changes in
  environmental dynamics.
\newblock Experimental Brain Research 2018;236(11):2923--2933.

\bibitem[{Brenner and Smeets(2003)Brenner, Eli and Smeets, Jeroen B.
  J.}]{brenner_fast_2003}
Brenner E, Smeets JBJ.
\newblock Fast corrections of movements with a computer mouse.
\newblock Spatial Vision 2003;16(3-4):365--376.

\bibitem[{Sarlegna et~al.(2003)Sarlegna, Fabrice and Blouin, Jean and
  Bresciani, Jean-Pierre and Bourdin, Christophe and Vercher, Jean-Louis and
  Gauthier, Gabriel M.}]{sarlegna_target_2003}
Sarlegna F, Blouin J, Bresciani JP, Bourdin C, Vercher JL, Gauthier GM.
\newblock Target and hand position information in the online control of
  goal-directed arm movements.
\newblock Experimental Brain Research 2003;151(4):524--535.

\bibitem[{Saunders and Knill(2003)Saunders, Jeffrey A. and Knill, David
  C.}]{saunders_humans_2003}
Saunders JA, Knill DC.
\newblock Humans use continuous visual feedback from the hand to control fast
  reaching movements.
\newblock Experimental Brain Research 2003;152(3):341--352.

\bibitem[{Franklin and Wolpert(2008)Franklin, David W. and Wolpert, Daniel
  M.}]{franklin_specificity_2008}
Franklin DW, Wolpert DM.
\newblock Specificity of reflex adaptation for task-relevant variability.
\newblock Journal of Neuroscience 2008;28(52):14165--14175.

\bibitem[{Knill et~al.(2011)Knill, David C. and Bondada, Amulya and Chhabra,
  Manu}]{knill_flexible_2011}
Knill DC, Bondada A, Chhabra M.
\newblock Flexible, task-dependent use of sensory feedback to control hand
  movements.
\newblock Journal of Neuroscience 2011;31(4):1219--1237.

\bibitem[{Oldfield(1971)Oldfield, R. C.}]{oldfield_assessment_1971}
Oldfield RC.
\newblock The assessment and analysis of handedness: the {Edinburgh} inventory.
\newblock Neuropsychologia 1971;9(1):97--113.

\bibitem[{Howard et~al.(2009)Howard, Ian S. and Ingram, James N. and Wolpert,
  Daniel M.}]{howard_modular_2009}
Howard IS, Ingram JN, Wolpert DM.
\newblock A modular planar robotic manipulandum with end-point torque control.
\newblock Journal of Neuroscience Methods 2009;181(2):199--211.

\bibitem[{Franklin et~al.(2014)Franklin, D. W. and Franklin, S. and Wolpert, D.
  M.}]{franklin_fractionation_2014}
Franklin DW, Franklin S, Wolpert DM.
\newblock Fractionation of the visuomotor feedback response to directions of
  movement and perturbation.
\newblock J Neurophysiol 2014;112(9):2218--2233.

\bibitem[{Dimitriou et~al.(2013)Dimitriou, M. and Wolpert, D. M. and Franklin,
  D. W.}]{dimitriou_temporal_2013}
Dimitriou M, Wolpert DM, Franklin DW.
\newblock The {Temporal} {Evolution} of {Feedback} {Gains} {Rapidly} {Update}
  to {Task} {Demands}.
\newblock Journal of Neuroscience 2013;33(26):10898--10909.

\bibitem[{Reichenbach et~al.(2014)Reichenbach, Alexandra and Franklin, David W.
  and Zatka-Haas, Peter and Diedrichsen, Jörn}]{reichenbach_dedicated_2014}
Reichenbach A, Franklin DW, Zatka-Haas P, Diedrichsen J.
\newblock A dedicated binding mechanism for the visual control of movement.
\newblock Current Biology 2014;24(7):780--785.

\bibitem[{Scheidt et~al.(2000)Scheidt, R. A. and Reinkensmeyer, D. J. and
  Conditt, M. A. and Rymer, W. Z. and Mussa-Ivaldi, F.
  A.}]{scheidt_persistence_2000}
Scheidt RA, Reinkensmeyer DJ, Conditt MA, Rymer WZ, Mussa-Ivaldi FA.
\newblock Persistence of motor adaptation during constrained, multi-joint, arm
  movements.
\newblock Journal of Neurophysiology 2000;84(2):853--862.

\bibitem[{{JASP Team}(2020)}]{JASP2020}
{JASP Team}, {JASP (Version 0.14.1)[Computer software]}; 2020.
\newblock \urlprefix\url{https://jasp-stats.org/}.

\bibitem[{Smith et~al.(2006)Smith, Maurice A. and Ghazizadeh, Ali and Shadmehr,
  Reza}]{smith_interacting_2006}
Smith MA, Ghazizadeh A, Shadmehr R.
\newblock Interacting adaptive processes with different timescales underlie
  short-term motor learning.
\newblock PLoS Biology 2006;4(6):e179.

\bibitem[{Howard et~al.(2012)Howard, Ian S. and Ingram, James N. and Franklin,
  David W. and Wolpert, Daniel M.}]{howard_gone_2012}
Howard IS, Ingram JN, Franklin DW, Wolpert DM.
\newblock Gone in 0.6 seconds: the encoding of motor memories depends on recent
  sensorimotor {States}.
\newblock Journal of Neuroscience 2012;32:12756--68.

\bibitem[{Gu et~al.(2016)Gu, C. and Wood, D. K. and Gribble, P. L. and Corneil,
  B. D.}]{gu_trial-by-trial_2016}
Gu C, Wood DK, Gribble PL, Corneil BD.
\newblock A {Trial}-by-{Trial} {Window} into {Sensorimotor} {Transformations}
  in the {Human} {Motor} {Periphery}.
\newblock Journal of Neuroscience 2016;36(31):8273--82.

\bibitem[{Cross et~al.(2019)Cross, K. P. and Cluff, T. and Takei, T. and Scott,
  S. H.}]{cross_visual_2019}
Cross KP, Cluff T, Takei T, Scott SH.
\newblock Visual {Feedback} {Processing} of the {Limb} {Involves} {Two}
  {Distinct} {Phases}.
\newblock Journal of Neuroscience 2019;39(34):6751--6765.

\bibitem[{Crevecoeur et~al.(2013)Crevecoeur, F. and Kurtzer, I. and Bourke, T.
  and Scott, S. H.}]{crevecoeur_feedback_2013}
Crevecoeur F, Kurtzer I, Bourke T, Scott SH.
\newblock Feedback responses rapidly scale with the urgency to correct for
  external perturbations.
\newblock Journal of Neurophysiology 2013;110(6):1323--1332.

\bibitem[{Česonis and Franklin(2020)Česonis, Justinas and Franklin, David
  W.}]{cesonis_time--target_2020}
Česonis J, Franklin DW.
\newblock Time-to-target explains task-dependent modulation of temporal
  feedback gain evolution.
\newblock eNeuro 2020;7:ENEURO.0514--19.2020.

\bibitem[{Hadjiosif and Smith(2015)Hadjiosif, Alkis M. and Smith, Maurice
  A.}]{hadjiosif_flexible_2015}
Hadjiosif AM, Smith MA.
\newblock Flexible control of safety margins for action based on environmental
  variability.
\newblock Journal of Neuroscience 2015;35(24):9106--9121.

\bibitem[{Wu et~al.(2014)Wu, Howard G. and Miyamoto, Yohsuke R. and Gonzalez
  Castro, Luis Nicolas and Ölveczky, Bence P. and Smith, Maurice
  A.}]{wu_temporal_2014}
Wu HG, Miyamoto YR, Gonzalez~Castro LN, Ölveczky BP, Smith MA.
\newblock Temporal structure of motor variability is dynamically regulated and
  predicts motor learning ability.
\newblock Nature Neuroscience 2014;17(2):312--321.

\bibitem[{Cluff et~al.(2019)Cluff, Tyler and Crevecoeur, Frederic and Scott,
  Stephen H.}]{Cluff_tradeoffs_2019}
Cluff T, Crevecoeur F, Scott SH.
\newblock Tradeoffs in optimal control capture patterns of human sensorimotor
  control and adaptation.
\newblock bioRxiv 2019;730713.

\bibitem[{Alhussein et~al.(2019)Alhussein, Laith and Hosseini, Eghbal A. and
  Nguyen, Katrina P. and Smith, Maurice A. and Joiner, Wilsaan
  M.}]{alhussein_dissociating_2019}
Alhussein L, Hosseini EA, Nguyen KP, Smith MA, Joiner WM.
\newblock Dissociating effects of error size, training duration, and amount of
  adaptation on the ability to retain motor memories.
\newblock Journal of Neurophysiology 2019;122(5):2027--2042.

\bibitem[{Fine and Thoroughman(2006)Fine, Michael S. and Thoroughman, Kurt
  A.}]{fine_motor_2006}
Fine MS, Thoroughman KA.
\newblock Motor adaptation to single force pulses: sensitive to direction but
  insensitive to within-movement pulse placement and magnitude.
\newblock Journal of Neurophysiology 2006;96(2):710--720.

\bibitem[{Wei and Körding(2009)Wei, Kunlin and Körding,
  Konrad}]{wei_relevance_2009}
Wei K, Körding K.
\newblock Relevance of {Error}: {What} {Drives} {Motor} {Adaptation}?
\newblock Journal of Neurophysiology 2009;101(2):655--664.

\bibitem[{Marko et~al.(2012)Marko, Mollie K. and Haith, Adrian M. and Harran,
  Michelle D. and Shadmehr, Reza}]{marko_sensitivity_2012}
Marko MK, Haith AM, Harran MD, Shadmehr R.
\newblock Sensitivity to prediction error in reach adaptation.
\newblock Journal of Neurophysiology 2012;108(6):1752--1763.

\bibitem[{Hayashi et~al.(2020)Hayashi, Takuji and Kato, Yutaro and Nozaki,
  Daichi}]{hayashi_divisively_2020}
Hayashi T, Kato Y, Nozaki D.
\newblock Divisively {Normalized} {Integration} of {Multisensory} {Error}
  {Information} {Develops} {Motor} {Memories} {Specific} to {Vision} and
  {Proprioception}.
\newblock Journal of Neuroscience 2020;40(7):1560--1570.

\bibitem[{Heald et~al.(2018)Heald, James B. and Franklin, David W. and Wolpert,
  Daniel M.}]{heald_increasing_2018}
Heald JB, Franklin DW, Wolpert DM.
\newblock Increasing muscle co-contraction speeds up internal model acquisition
  during dynamic motor learning.
\newblock Scientific Reports 2018;8(1):16355.

\bibitem[{Gonzalez~Castro et~al.(2011)Gonzalez Castro, Luis Nicolas and Monsen,
  Craig Bryant and Smith, Maurice A.}]{gonzalez_castro_binding_2011}
Gonzalez~Castro LN, Monsen CB, Smith MA.
\newblock The binding of learning to action in motor adaptation.
\newblock PLoS Computational Biology 2011;7(6):e1002052.

\bibitem[{Huang et~al.(2012)Huang, Helen J. and Kram, Rodger and Ahmed, Alaa
  A.}]{huang_reduction_2012}
Huang HJ, Kram R, Ahmed AA.
\newblock Reduction of metabolic cost during motor learning of arm reaching
  dynamics.
\newblock Journal of Neuroscience 2012;32(6):2182--2190.

\bibitem[{Huang and Ahmed(2014)Huang, Helen J. and Ahmed, Alaa
  A.}]{huang_reductions_2014}
Huang HJ, Ahmed AA.
\newblock Reductions in muscle coactivation and metabolic cost during
  visuomotor adaptation.
\newblock Journal of Neurophysiology 2014;112(9):2264--2274.

\bibitem[{Darainy and Ostry(2008)Darainy, Mohammad and Ostry, David
  J.}]{darainy_muscle_2008}
Darainy M, Ostry DJ.
\newblock Muscle cocontraction following dynamics learning.
\newblock Experimental Brain Research 2008;190(2):153--163.

\bibitem[{Berniker and Körding(2008)Berniker, Max and Körding, Konrad
  P.}]{berniker_estimating_2008}
Berniker M, Körding KP.
\newblock Estimating the sources of motor errors for adaptation and
  generalization.
\newblock Nature Neuroscience 2008;11:1454--61.

\bibitem[{Crevecoeur et~al.(2010)Crevecoeur, F. and McIntyre, J. and Thonnard,
  J.-L. and Lefèvre, P.}]{crevecoeur_movement_2010}
Crevecoeur F, McIntyre J, Thonnard JL, Lefèvre P.
\newblock Movement stability under uncertain internal models of dynamics.
\newblock Journal of Neurophysiology 2010;104(3):1301--1313.

\bibitem[{Burdet et~al.(2006)Burdet, E. and Tee, K. P. and Mareels, I. and
  Milner, T. E. and Chew, C. M. and Franklin, D. W. and Osu, R. and Kawato,
  M.}]{burdet_stability_2006}
Burdet E, Tee KP, Mareels I, Milner TE, Chew CM, Franklin DW, et~al.
\newblock Stability and motor adaptation in human arm movements.
\newblock Biological Cybernetics 2006;94(1):20--32.

\bibitem[{McIntyre et~al.(1996)McIntyre, J. and Mussa-Ivaldi, F. A. and Bizzi,
  E.}]{mcintyre_control_1996}
McIntyre J, Mussa-Ivaldi FA, Bizzi E.
\newblock The control of stable postures in the multijoint arm.
\newblock Experimental Brain Research 1996;110(2):248--264.

\bibitem[{Franklin and Milner(2003)Franklin, David W. and Milner, Theodore
  E.}]{franklin_adaptive_2003}
Franklin DW, Milner TE.
\newblock Adaptive control of stiffness to stabilize hand position with large
  loads.
\newblock Experimental Brain Research 2003;152(2):211--220.

\bibitem[{Kawato et~al.(1987)Kawato, M. and Furukawa, K. and Suzuki,
  R.}]{kawato_hierarchical_1987}
Kawato M, Furukawa K, Suzuki R.
\newblock A hierarchical neural-network model for control and learning of
  voluntary movement.
\newblock Biological Cybernetics 1987;57(3):169--185.

\bibitem[{Franklin et~al.(2008)Franklin, David W. and Burdet, Etienne and Tee,
  Keng Peng and Osu, Rieko and Chew, Chee-Meng and Milner, Theodore E. and
  Kawato, Mitsuo}]{franklin_cns_2008}
Franklin DW, Burdet E, Tee KP, Osu R, Chew CM, Milner TE, et~al.
\newblock {CNS} learns stable, accurate, and efficient movements using a simple
  algorithm.
\newblock Journal of Neuroscience 2008;28(44):11165--11173.

\bibitem[{Albert and Shadmehr(2016)Albert, S. T. and Shadmehr,
  R.}]{albert_neural_2016}
Albert ST, Shadmehr R.
\newblock The {Neural} {Feedback} {Response} to {Error} {As} a {Teaching}
  {Signal} for the {Motor} {Learning} {System}.
\newblock Journal of Neuroscience 2016;36(17):4832--4845.

\bibitem[{Patton and Mussa-Ivaldi(2004)Patton, James L. and Mussa-Ivaldi,
  Ferdinando A.}]{patton_robot-assisted_2004}
Patton JL, Mussa-Ivaldi FA.
\newblock Robot-assisted adaptive training: custom force fields for teaching
  movement patterns.
\newblock IEEE Transactions on Biomedical Engineering 2004;51(4):636--646.

\bibitem[{Huang and Patton(2012)Huang, F. and Patton,
  J.}]{huang_augmented_2012}
Huang F, Patton J.
\newblock Augmented dynamics and motor exploration as training for stroke.
\newblock IEEE Transactions on Biomedical Engineering 2012;60(3):838--844.

\bibitem[{Reinkensmeyer et~al.(2016)Reinkensmeyer, David J. and Burdet, Etienne
  and Casadio, Maura and Krakauer, John W. and Kwakkel, Gert and Lang,
  Catherine E. and Swinnen, Stephan P. and Ward, Nick S. and Schweighofer,
  Nicolas}]{reinkensmeyer_computational_2016}
Reinkensmeyer DJ, Burdet E, Casadio M, Krakauer JW, Kwakkel G, Lang CE, et~al.
\newblock Computational neurorehabilitation: modeling plasticity and learning
  to predict recovery.
\newblock Journal of Neuroengineering and Rehabilitation 2016;13(1):1--26.

\bibitem[{Torres-Oviedo and Bastian(2012)Torres-Oviedo, Gelsy and Bastian, Amy
  J.}]{torres-oviedo_natural_2012}
Torres-Oviedo G, Bastian AJ.
\newblock Natural error patterns enable transfer of motor learning to novel
  contexts.
\newblock Journal of Neurophysiology 2012;107(1):346--356.

\bibitem[{Pruszynski et~al.(2009)Pruszynski, J. Andrew and Kurtzer, Isaac and
  Lillicrap, Timothy P. and Scott, Stephen H.}]{pruszynski_temporal_2009}
Pruszynski JA, Kurtzer I, Lillicrap TP, Scott SH.
\newblock Temporal evolution of "automatic gain-scaling".
\newblock Journal of Neurophysiology 2009;102(2):992--1003.

\bibitem[{Joiner et~al.(2011)Joiner, Wilsaan M. and Ajayi, Obafunso and Sing,
  Gary C. and Smith, Maurice A.}]{joiner_linear_2011}
Joiner WM, Ajayi O, Sing GC, Smith MA.
\newblock Linear hypergeneralization of learned dynamics across movement speeds
  reveals anisotropic, gain-encoding primitives for motor adaptation.
\newblock Journal of Neurophysiology 2011;105(1):45--59.

\bibitem[{Franklin et~al.(2007)Franklin, David W. and Liaw, Gary and Milner,
  Theodore E. and Osu, Rieko and Burdet, Etienne and Kawato,
  Mitsuo}]{franklin_endpoint_2007}
Franklin DW, Liaw G, Milner TE, Osu R, Burdet E, Kawato M.
\newblock Endpoint stiffness of the arm is directionally tuned to instability
  in the environment.
\newblock Journal of Neuroscience 2007;27(29):7705--7716.

\bibitem[{Wagner and Smith(2008)Wagner, Mark J. and Smith, Maurice
  A.}]{wagner_shared_2008}
Wagner MJ, Smith MA.
\newblock Shared internal models for feedforward and feedback control.
\newblock Journal of Neuroscience 2008;28(42):10663--10673.

\bibitem[{Ahmadi-Pajouh et~al.(2012)Ahmadi-Pajouh, Mohammad Ali and Towhidkhah,
  Farzad and Shadmehr, Reza}]{ahmadi-pajouh_preparing_2012}
Ahmadi-Pajouh MA, Towhidkhah F, Shadmehr R.
\newblock Preparing to reach: selecting an adaptive long-latency feedback
  controller.
\newblock Journal of Neuroscience 2012;32(28):9537--9545.

\bibitem[{Maeda et~al.(2018)Maeda, Rodrigo S. and Cluff, Tyler and Gribble,
  Paul L. and Pruszynski, J. Andrew}]{maeda_feedforward_2018}
Maeda RS, Cluff T, Gribble PL, Pruszynski JA.
\newblock Feedforward and feedback control share an internal model of the arm's
  dynamics.
\newblock Journal of Neuroscience 2018;38(49):10505--10514.

\bibitem[{Malfait et~al.(2002)Malfait, Nicole and Shiller, Douglas M. and
  Ostry, David J.}]{malfait_transfer_2002}
Malfait N, Shiller DM, Ostry DJ.
\newblock Transfer of motor learning across arm configurations.
\newblock Journal of Neuroscience 2002;22(22):9656--9660.

\bibitem[{Berniker et~al.(2014)Berniker, Max and Franklin, David W. and
  Flanagan, J. Randall and Wolpert, Daniel M. and Kording,
  Konrad}]{berniker_motor_2014}
Berniker M, Franklin DW, Flanagan JR, Wolpert DM, Kording K.
\newblock Motor learning of novel dynamics is not represented in a single
  global coordinate system: evaluation of mixed coordinate representations and
  local learning.
\newblock Journal of Neurophysiology 2014;111(6):1165--1182.

\bibitem[{Ueyama(2014)Ueyama, Yuki}]{ueyama_mini-max_2014}
Ueyama Y.
\newblock Mini-max feedback control as a computational theory of sensorimotor
  control in the presence of structural uncertainty.
\newblock Frontiers in Computational Neuroscience 2014;8:119.

\bibitem[{Crevecoeur et~al.(2019)Crevecoeur, Frédéric and Scott, Stephen H.
  and Cluff, Tyler}]{crevecoeur_robust_2019}
Crevecoeur F, Scott SH, Cluff T.
\newblock Robust {Control} in {Human} {Reaching} {Movements}: {A}
  {Model}-{Free} {Strategy} to {Compensate} for {Unpredictable} {Disturbances}.
\newblock Journal of Neuroscience 2019;39(41):8135.

\bibitem[{Bian et~al.(2020)Bian, Tao and Wolpert, Daniel M. and Jiang,
  Zhong-Ping}]{bian_model-free_2020}
Bian T, Wolpert DM, Jiang ZP.
\newblock Model-{Free} {Robust} {Optimal} {Feedback} {Mechanisms} of
  {Biological} {Motor} {Control}.
\newblock Neural Computation 2020;32(3):562--595.

\end{thebibliography}



\end{document}